\documentclass[12pt,preprint]{aastex}

\usepackage{graphics,graphicx}
\usepackage{amsmath}
\usepackage{natbib}
\usepackage{color}
\def\vs{{\mathbf{J}}}

\newcommand{\gtorder}{\mathrel{\raise.3ex\hbox{$>$}\mkern-14mu
            \lower0.6ex\hbox{$\sim$}}}
\newcommand{\ltorder}{\mathrel{\raise.3ex\hbox{$<$}\mkern-14mu
            \lower0.6ex\hbox{$\sim$}}}

\shorttitle{Sound Speed and Tilt Dependence}
\shortauthors{Hawley \& Krolik}

\begin{document}

    \title{Tilt Dependence of Alignment in Accretion Disks Subjected to Lense--Thirring Torques}

\author{John F. Hawley\altaffilmark{1}}

\and

\author{Julian H. Krolik\altaffilmark{2}}

\altaffiltext{1}{Department of Astronomy, University of Virginia, Charlottesville VA 22904, USA 0000-0002-0376-0318}

\altaffiltext{2}{Department of Physics and Astronomy, Johns Hopkins University, Baltimore, MD 21218, USA 0000-0002-2995-7717)} 

\begin{abstract}
We consider the effects of  black hole tilt on accretion disk alignment,  studying three initial black hole tilts, $6^\circ$, $12^\circ$ and  $24^\circ$, with both magnetohydrodynamic and (inviscid) hydrodynamic evolution.   In a number of ways, but not all, the dynamics are homologous in the sense that the alignment fronts resulting from different initial tilts are very similar when analyzed in terms of the fraction of the initial tilt angle.  Even when the initial misalignment is $24^\circ$, which, for the sound speed studied, is 4 vertical scale heights at the disk fiducial radius, the surface density remains a smooth function of radius; i.e., we find no examples in which the disk  inner aligned and outer misaligned regions separate, or ``break".
\end{abstract}


\section{Introduction}

As a simple combination of gravity and angular momentum, disks are common astrophysical systems.  When the gravitating object's angular momentum does not align with the orbiting matter's angular momentum, complex dynamics result due to precessional torques.  A disk within a binary system that is misaligned with the plane of the binary, for example, is subject to torque due to the binary's quadrupole potential.  A closely analogous torque, the Lense--Thirring torque, arises when a disk around a single black hole is misaligned with the black hole's spin.  The question in either case is how the disk evolves under the influence of that torque.  Since the work of \cite{BP75}, it has been expected that in the latter case the disk comes into alignment in the inner region where the torque is relatively strong, but retains its original orientation at large radii where the torque is negligible.  More than forty years later much about this scenario remains uncertain, particularly about the transition region between these two limits.

Both quadrupole and Lense--Thirring torques produce precession, but not alignment, because the torque is exactly perpendicular to the local angular momentum.   Alignment, therefore, must be due to the disk acquiring angular momentum from some other place where the local angular momentum has a different direction, and this new angular momentum must then be transported to a location where, when added to the local angular momentum, the disk is rotated toward alignment.  For this to happen, the precession phase where the external torque acts must be advanced relative to the place where the angular momentum is ultimately delivered.   Thus, the mechanics of alignment are all about internal angular momentum transport.

The transport of angular momentum within a disk is, of course, the central question of accretion physics, regardless of whether or not the disk is aligned \citep[see the review by][on disk angular momentum transport processes]{Balbus2003}.  In thin, flat disks, where radial pressure gradients are very small, the mechanism that transports orbital angular momentum in the radial direction, i.e., the nature of the $r$--$\phi$ component of the internal stress, $t_{r\phi}$, remained obscure for many years.  Early in the development of the field, it was estimated on the basis of dimensional analysis: $t_{r\phi} = \alpha P$, where $P$ is the local thermal (gas plus radiation) pressure \citep{SS73}.  Twenty years later, now more than twenty-five years ago, a robust physical mechanism for this stress was found: stirring of magnetohydrodynamic (MHD) turbulence by the magneto-rotational instability \citep[MRI;][]{mri91,BH98}.  Although much remains uncertain about what exactly determines the saturation amplitude of this turbulence, and therefore the magnitude of the associated $t_{r\phi}$, two things are clear.  One is that the mean value of $t_{r\phi} \neq 0$ because the consistent sense of orbital shear imposes a consistent asymmetry on the turbulence.  The other is that the long--term average of the vertically integrated stress measured well away from any radial boundaries of the disk does appear to be roughly proportional to the similarly averaged vertical pressure.

Aligning angular momentum is carried by a different component of the stress, $t_{rz}$.  A physical mechanism to create such a stress was identified early on \citep{PP83}.  Precessional torques generically decline rapidly in strength with increasing radius, thereby inducing differential precession.   Differential precession creates disk warps, disk warps create radial pressure gradients, and radial pressure gradients induce radial motions.  The net result is a non-zero Reynolds stress $t_{rz}$ derived from bulk radial motions, rather than from a microscopic diffusive mechanism. In the decades since, there has been much controversy over how this stress's magnitude is regulated.

From its inception, quantitatve exploration of this stress's origins and consequences has largely focused on analytical approaches \citep[e.g.,][]{BP75,Hatchett81,PP83,Pringle92,PapLin95,II97,O99,Lubow02}.  Since the work of \citet{PP83}, this research program has assumed that the $\alpha$ prescription for the $t_{r\phi}$ stress can be generalized to predict the $t_{rz}$ stress by supposing that in response to shear in the radial fluid motions, a viscous stress is created whose magnitude is $\alpha P$ times the shear in units of the orbital shear.  Although a bold speculation, this {\it ansatz} has proven very attractive because it provides a way to proceed in the absence of an understanding of internal disk stresses and also leads to well-defined equations amenable to solution.

On the other hand, it remains unphysical.  Moreover, thanks to the MRI paradigm, we actually do understand the origin of the $t_{r\phi}$ stress, and it is not intrinsically viscous \citep{BH98,Balbus2003,Blaes2014}.  Nor is there any reason to think that magnetorotational effects should sustain $t_{rz}$ stress--- orbital shear contributes to the $r$-$\phi$ component of the shear tensor, not the $r$-$z$ component.

The supposition of ``isotropic $\alpha$ viscosity" also led to the suggestion that there are two different regimes of warp evolution, depending on whether $\alpha$ is greater or smaller than the disk's ratio of vertical scale height $h$ to local radius $r$ \citep{PP83,PapLin95}.  In the former case, warp evolution is deemed ``diffusive", while in the latter it is supposed to be ``bending wave dominated".  However, the unphysicality of ``$\alpha$ viscosity" also throws this categorization into question.

Motivated by these considerations, for the last five years we have pursued an alternative approach in our investigation of alignment: focused numerical experiments employing MHD simulations. It has also been instructive to contrast these simulations with paired simulations of inviscid, purely hydrodynamic (HD) disks.

Several points have been demonstrated by these simulations.  First, in purely hydrodynamic disks, warps evolve due to the propagation of bending waves, and the nature of their propagation depends strongly on their amplitude, but not particularly on the level of accretion stress (i.e., the $\alpha$ parameter, which is zero in the case of purely HD disks).  In \citet{SKH13a}, we explored quantitatively a point first made by \citet{NP99}: when $\hat\psi \equiv | d\hat\ell/d\ln r|/(h/r) > 1$, where $\hat\ell$ is the unit vector in the direction of a disk ring's angular momentum, bending waves may be considered ``nonlinear" in the sense that the radial pressure contrast they induce is comparable to the baseline disk pressure itself.  Such nonlinear waves drive shocks, and dissipation in these shocks rapidly damps the wave amplitude.  In addition, the rate at which such warps are smoothed is not particularly well described by Fick's Law: although the stress induced does increase with normalized warp amplitude $\hat\psi$, the dependence is not linear, and there are time-delays of order the local orbital period.  Thus, time-dependent warp dynamics cannot be modeled accurately with diffusion equations.

Second, when magnetic fields are introduced in the context of a misaligned disk subjected to external precessional torques \citep{SKH13b}, the internal stress associated with the induced warps can be directly measured.  It was found to be several orders of magnitude smaller than predicted by the ``isotropic $\alpha$ viscosity" model, and, more tellingly, its {\it sign} was as likely to be with the shear as against it, demonstrating that it cannot be thought of as any sort of viscosity.  Although MHD turbulence dominates the generation of $t_{r\phi}$ stress, its primary role in alignment appears to be in disrupting the global communication within the disk (mediated by bending waves) that leads to solid body precession, the onset of which prevents further alignment.  Thus, we have found that regulation of the $t_{rz}$ stress is governed by inviscid hydrodynamics (shocks, etc.), while the propagation of bending waves is controlled by turbulent perturbations in the background medium through which they travel.  In neither context is there a role for a putative viscosity.

In the third portion of this program, we focused on the properties of steady state alignment fronts \citep{KH2015,HK18}.  Although diffusion does not describe warp evolution very well, it does seem to be a reasonable approximation for time-steady warp properties.  Using MHD and HD simulations spanning a range of values of $h/r$, \citet{HK18} showed that a simple ``lumped parameter'' diffusion model for the steady state transition radius between aligned and unaligned portions of the disk is quantitatively accurate to the point that the simulations calibrate the dimensionless coefficient of the warp diffusion coefficient, which, by dimensional analysis, must be $\propto c_s^2/\Omega$, for sound speed $c_s$ and orbital frequency $\Omega$.  Although the ratio of accretion stress to disk aspect ratio spanned by our simulations corresponded to values of $\alpha/(h/r)$ ranging from $\approx 0.1$ to $\approx 4$, they differed very little in their development, further undermining the $\alpha$-based ``diffusive regime" vs. ``bending wave regime" dichotomy.

Another controversy arises from the possible dependence of warped disk mechanics on the amplitude of the intrinsic misalignment.  Most dramatically, there have been reports that disks may ``break" if the intrinsic tilt is too large.
The first hints came in the work of \citet{Larwood96} and \citet{NP00}.  In the former, smooth-particle hydrodynamic (SPH) simulations of a binary system containing a disk with $h/r < 0.03$ and $\theta = 45^\circ$ suggested that the disk's aligned inner portion was separating from its unaligned outer portion.  In the latter, SPH simulations of comparably cool disks tilted by $30^\circ$ relative to a black hole exerting Lense--Thirring torque came ``close to breaking".  Further hints appeared in \citet{Lodato10}, in which a strongly warped disk ($\hat\psi \sim 100$) with an $\alpha$ viscosity, but no external torque, developed a gradual steepening of the transition zone leading to a disconnect between the inner and outer disk.

These hints were further developed by \citet{Nixon2012a}, who modeled a thin disk misaligned with a spinning black hole and subject to three effective viscosities.  They found that in this model system, disk breaking can occur for sufficiently large warp amplitudes (i.e., sufficiently large $|\partial \hat \ell/\partial \ln r|$) or when the radially-acting viscosity is small enough to render the disk unable to communicate in the radial direction.
\cite{Nixon2012b}, utilizing the results of both order-of-magnitude analysis and SPH simulations of disks with $\alpha$ viscosity and tilts as large as $60^\circ$, proposed that disk breaking occurs where the Lense--Thirring torque exceeds the internal viscous torque. This supposition predicts
breaks can be expected when $|\sin \theta | \ge (3/4)(\alpha/a_*)(h/r)$, where $a_*$ is the black hole spin parameter.  If so, breaks would be almost ubiquitous, as the minimum tilt would be quite small unless the black hole hardly spins.
\cite{Nealon2015} performed further SPH simulations of the Bardeen-Petterson effect,
considering a wide range of alignment angles and black hole spin parameters; they found disk breaks in almost all cases where the numerical resolution was considered adequate, including multiple breaks when the tilt angle was large.  However, in a number of cases the break location did not match any of the scaling argument predictions, neither the one obtained by equating internal and external torques nor the one derived from equating the sound-crossing and precession times.  Rather, the break radius
was nearly constant as a function of $a_*$.  Thus, the current situation with respect to the existence of breaks and the behavior of large-tilt disks in general remains unclear.

In this paper, we seek to explore the effect of tilt angle without any parameterized viscosity, employing instead well-resolved MHD simulations and contrasting them with pure (i.e., no artificial viscosity) HD simulations.

\section{Simulations}

\subsection{Model system and numerics}

We use the model system first studied in \cite{KH2015} and subsequently in \cite{HK18}, which is an idealization designed for detailed investigations of the alignment mechanism.  The model  consists of an isothermal disk orbiting a point-mass in Newtonian gravity with a Keplerian angular velocity distribution, $\Omega^2 = GM/r^3$.  Since we employ a Newtonian potential, the radial units are arbitrary, in contrast to both relativistic gravity or a pseudo-Newtonian potential defined in terms of a gravitational radius $r_g = GM/c^2$.    The Lense--Thirring effect is included through the addition of the lowest-order post Newtonian term.   We set $GM = 1$, so that the Lense--Thirring precession frequency $\Omega_{\rm precess} = 2/r^3$ if $a_*=1$.  This frequency is $\Omega(r_*)/15.8$ at the fiducial radius $r_*=10$.  We report time in units of fiducial orbits, defined as 200 units of code time, which is almost exactly the orbital period  $P_{\rm orb} = 2\pi r^{3/2} = 199$ at $r=10$.
As in the previous work, we use our Fortran-95 implementation of the 3D finite-difference algorithm known as {\it Zeus} \citep{zeus1,zeus2,hawleystone95}.  The {\it Zeus} code solves the standard equations of Newtonian MHD (supplemented by the torque term previously described) using direct finite differencing.  We use spherical coordinates $(r,\theta,\phi)$. 

We consider an isothermal disk model with sound speed $c_s^2 = 2.5\times 10^{-4}$, which is the same sound speed as the ``High-thin'' model of \cite{HK18}.  This sound speed gives $h/r = 0.05$ at the fiducial radius.  
The density at the equator of the disk is $\rho_c = 1$ at all radii, and the vertical distribution is set by by assuming vertical hydrostatic equilibrium, i.e., $\rho = \rho_c \exp (-z^2/2h^2)$.   At the initial inner ($r=6$) and outer disk limits, the disk is truncated.  Consequently, the disk is not in radial pressure equilibrium at the disk boundaries, and in the subsequent evolution the disk's outer boundary moves outward from where the disk was initially truncated.
The surface density $\Sigma \propto h$ increases outward $\propto r^{3/2}$ until the outer portion of the disk where, due to the finite size of the disk, $\Sigma$ smoothly declines to zero. 

We add a ``seed'' initial magnetic field defined by a vector potential proportional to the square root of the disk density within an ``envelope'' function,
\begin{equation}
A_{\phi} = A_0 \rho^{1/2} \sin \left[{\pi\over 2} ({r_o/ r})^{1/2}\right] (r/r_{in} -1) (1-r/r_{out}) 
\end{equation}
where $r_o =4$ is the grid inner boundary, $r_{in}$ is the disk inner radius and $r_{out}$ is the disk outer radius.
The vector potential is limited to positive values with a cutoff at $0.05\rho_c$, i.e.,
\begin{equation}
A_{\phi} = \max(A_{\phi} - 0.05 \rho_{c},0).
\end{equation}  
The field amplitude factor $A_0$ is chosen so that the initial volume-integrated ratio of gas to magnetic pressure, the plasma $\beta$, is 1000.  This particular vector potential leads to weak, primarily radial,  magnetic field that rapidly generates toroidal field through Keplerian shear.  This field is subject to the MRI, which leads to turbulence and non-zero $t_{r\phi}$ as the field amplitude grows.

Using this isothermal disk model, we study the effect of the black hole tilt angle on alignment.  
The ``High-thin'' model of \cite{HK18} was evolved with a tilt angle of $12^\circ$.  In this paper we shall refer to this model as ``Tilt12.''  Here we also examine disk behavior when the tilt angle is doubled to $24^\circ$ with a simulation called Tilt24, as well as one with the tilt angle halved to $6^\circ$, referred to as Tilt6.

Table~\ref{table:list}  lists the models and their parameters.
The table gives: the name of the simulation; the number of grid-cells employed; the sound speed;  the run duration with torque in units of fiducial orbits; and the radius of the outer boundary of the disk at the onset of the Lense--Thirring torque.  This last quantity is defined as the radius where the azimuthally-averaged surface density drops below 5\% of the initial maximum value.

\begin{deluxetable}{lccccc}
\tabletypesize{\scriptsize}
\tablewidth{0pc}
\tablecaption{Simulation List\label{table:list}}
\tablehead{
  \colhead{Name}&
  \colhead{$(r,\theta,\phi)$}&
  \colhead{$c_s^2$}&
  \colhead{Orbits}&
  \colhead{Tilt Angle}&
  \colhead{$r_{\rm out}$} 
  }
\startdata
Tilt6         & $714\times 768\times 1024$    &$2.5\times 10^{-4}$& 19.5  & $6^\circ$ & 28   \\
Tilt6-H      & $320\times 400\times 500$    &$2.5\times 10^{-4}$& 27.3  & $6^\circ$ & 28   \\
Tilt12    & $704\times 770\times 1024$  &$2.5\times 10^{-4}$&   22.3 & $12^\circ$ &  31 \\
Tilt12-H       & $320\times 400\times 500$    &$2.5\times 10^{-4}$&  33.8 & $12^\circ$ & 28   \\
Tilt24          & $960\times 1024\times 1024$&$2.5\times 10^{-4}$& 20.5 & $24^\circ$ & 33 \\
Tilt24-H      & $400\times 400\times 400$    &$2.5\times 10^{-4}$& 18.5&$24^\circ$&  40
\enddata
\end{deluxetable}

All simulations use similar spherical grids, but with different resolutions and spacings as needed.   In all three, the radial grid extends outward from a minimum value using a logarithmically graded mesh.  Because we are working with relatively thin disks, and we wish to avoid potential difficulties with coordinate singularities near the axis, we limit the extent of $\theta$ to the interval $[0.1,0.9]\pi$.  \cite{Sorathia12} showed that  increased numerical dissipation is associated
with increased orbital motion across the $(\theta, \phi)$ sphere compared to orbiting purely in the $\phi$ direction.  Therefore, to limit the potential increase in numerical dissipation, we use a large number of $\theta$ zones and concentrate them around the equatorial plane using the polynomial spacing given by equation (6) of  \cite{NKH10}, 
\begin{equation}\label{eqn:grid}
\theta(y) = {\pi \over 2}\left[ 1+(1-\xi)(2y-1) + (\xi - {2\theta_c \over \pi})(2y-1)^n \right]
\end{equation}
The $\theta$ grid index is $y= (i+0.5)/N$, where $i$ is the zone-index and $N$ is the total number of $\theta$ zones.  The angle $\theta_c$ gives the size of the ``cutout'' around the polar grid axis; this is $0.1\pi$ for all three models.  For the Tilt12 \citep{HK18} and Tilt6 models $\xi = 0.65$, and $n = 13$. The resulting distribution of zones has a relatively large $\Delta \theta$ near the cutouts along the axis, but the cell-size smoothly decreases to a small, constant $\Delta \theta$ over a symmetrical region surrounding the equator.  Tilt24 requires an increase in the number of  $\theta$ zones to cover the wider polar angle through which the disk moves, as well as a larger region over which the smallest $\Delta\theta$ zones are employed.  Thus, in Tilt24 we both increase the total number of $\theta$ zones and set $\xi=0.3$, and $n=13$. 
The $\phi$ coordinate covers the full $2\pi$ in angle with uniform spacing. Outflow boundary conditions are employed on the radial inner and outer boundaries, and along the $\theta$ boundary that forms a ``cut-out'' around the grid polar axis.

We establish Cartesian coordinates to describe how the disk tilts and warps, choosing the direction of the black hole spin $\vs$ to define the $z$ axis.  The polar axis of the code's spherical grid, which is parallel to the initial disk angular momentum, is in the $x$-$z$ plane.  At each radius we compute a shell average of the disk angular momentum, $\vec \ell (r)$, and transform the resulting averaged vector into the Cartesian system.  From this we compute the alignment angle
\begin{equation}\label{eq:beta}
\beta = \tan^{-1}\left(|\ell_\perp|/|\ell_z|\right),
\end{equation}
where $\ell_\perp^2 = \ell_x^2+\ell_y^2$, the precession angle 
\begin{equation}\label{eq:phi}
\phi_{prec} = \tan^{-1}\left(\ell_y/\ell_x\right),
\end{equation}
which runs from 0 to $2\pi$ as $\hat\ell$ precesses around the $z$ axis, 
and the total warp rate $\psi \equiv |\partial \hat \ell/\partial \ln r|$.  We define  $\hat\psi$ to be the warp rate normalized to the local value of $h/r$, i.e., $\hat\psi = \psi/(h/r)$.

 To establish an MHD turbulent disk, the models were initially computed without
any applied Lense--Thirring torque.  At the end of the initial ``no torque'' phase, the ``MRI quality metrics'' \citep{HRGK13} were measured.  The ``Q'' metrics for each coordinate axis measure the average number of grid cells across the fastest growing MRI mode whose wavevector is in that coordinate direction; these wavelengths are proportional to the magnetic field component for the corresponding axis.   

Tilt12 was evolved for 18 orbits without torque. At the end of this time, the value of $Q_\phi$  increases with radius from $\sim 10$ at $r=5$ to $>30$ for $r > 9$.  $Q_\theta$ has a similar profile, but is only $\approx 0.85 Q_\phi$.  These values indicate that the primary MRI wavelengths are well-resolved. 
At the same time, $\alpha_{mag}$, which is the ratio of the magnetic stress to the magnetic pressure, had an average value of 0.28 and $\langle B_r^2\rangle/\langle B_\phi^2 \rangle$ had an average value of 0.14 between $r=5$ and $25$ (the main portion of the disk).   
These quality metric values are somewhat below the values associated observed with well-developed MHD turbulence in highly resolved shearing sheet simulations, namely $\alpha_{\rm mag} \sim 0.4$ and $\langle B_r^2\rangle/\langle B_\phi^2\rangle\sim 0.2$ \citep{HGK11}. 

Tilt6 was evolved for 15 orbits without torque to allow the MHD turbulence to develop.  At the end of this initial evolution the average quality values between $r=6$ and 20 were $Q_\phi = 33$, $Q_\theta = 27$, $\alpha_{\rm mag} = 0.27$, and $\langle B_r^2\rangle/\langle B_\phi^2\rangle = 0.14 $.  

To reduce the computational time required, the Tilt24 initial disk without torque was run on a reduced $\phi$ domain of $\pi/4$ for 16.5 orbits, after which it was mapped onto the full $2\pi$ domain, perturbations added (to break the $m=4$ symmetry), and torque turned on.
 At this time the average quality values between $r=6$ and 20 were $Q_\phi = 37$, $Q_\theta = 21$, $\alpha_{\rm mag} = 0.29$, and $\langle B_r^2\rangle/\langle B_\phi^2\rangle = 0.15 $.  
 
For each of these MHD simulations we computed a paired purely HD model; these are designated by appending ``-H'' to the simulation name.  Because there is no need to resolve the MRI or the resulting turbulence, the HD simulations were run at lower resolution.  Tilt12-H was computed in \cite{HK18}; Tilt6-H used the same grid and initial hydro disk as that model.  Tilt24-H used a new initial disk which was evolved in axisymmetry for 20 orbits of time to allow for the initial transients to settle down.  At this point the disk was mapped onto the full $2\pi$ in $\phi$, perturbations were added and the torque applied. 

\section{Results}
\subsection{Expectations from previous work}

We begin with a brief review of relevant past results to set the stage for this study of the impact of tilt angle on aligment. \cite{SKH13b} and \cite{KH2015}  proposed that, in the absence of any mechanism to mix in misaligned angular momentum from larger radii, the propagation speed of the alignment front is determined by the rate at which angular momentum whose direction could cancel the misalignment could be carried outward in the disk.  This rate is characterized by the angle $\gamma$ between the angular momentum (perpendicular to the black hole spin axis) being carried outward and the direction opposite to the local misaligned angular momentum (here we use ``local" to mean ``averaged on a spherical shell").  The transported angular momentum optimally cancels the misaligned angular momentum when $\gamma = 0$.  The local torque scales with the surface density $\Sigma$ and $\sin\beta$, the local misalignment.  The alignment front propagation speed is the ratio of this torque to the local misaligned angular momentum, and is given by
\begin{equation}\label{eq:rfspeed2}
{d r_f\over dt} = \langle \cos\gamma\rangle \mathcal {I}(r_f) r_f \Omega_{\rm precess},
\end{equation}
where $r_f$ is the location of the head of the alignment front, and the averaging over $\cos\gamma$ refers to an average over the turbulence.
${\mathcal I}$ is the dimensionless integral
\begin{equation}\label{eq:scrI}
\mathcal{I}(r)   = \int_0^1 \, dx \, x^{-3/2} \frac{\sin\beta(x)}{\sin\beta(r_f)}
   \frac{\Sigma(x)}{\Sigma(r_f)},
\end{equation}
in which $x = r/r_f$.
This outward propagation is ultimately limited by the radial mixing induced by the radial pressure gradients associated with the disk's warp.  Comparison with simulation data \citep{HK18} has shown that time-steady properties due to this mixing (but {\it not} time-dependent ones) can be described by a diffusion model.  Although our expression for the time-steady alignment front location is
based on a linear diffusion equation for warp evolution, information
from our fully nonlinear simulations is embedded in this equation
through the value of the ``diffusion" coefficient.   The rate at which
misaligned angular momentum is mixed radially inward, and therefore the
value of this coefficient, is determined by highly nonlinear
hydrodynamics.
Equating the alignment front speed with the diffusion velocity yields
\begin{equation}
\langle -\cos\gamma\rangle {\mathcal I} r_T \Omega_{\rm precess} = A \left[c_s^2 /(r_T \Omega)\right] B(r_T),
\end{equation}
where $\Omega$ is the local orbital frequency  and $A$ is the dimensionless factor in the diffusion coefficient.  The quantity $B \equiv  |\partial \sin\beta/\partial \ln r| / \sin\beta$ calibrates the rate at which misaligned angular momentum is transported radially by diffusion.
Inserting the radial dependences for $\Omega_{\rm precess}$ and the orbital frequency, we find that the steady-state alignment front $r_T$ is located at
\begin{equation}\label{eq:rT}
r_{T}/r_g = \left[\frac{ 2(a/M) \langle \cos\gamma\rangle {\mathcal I}}{AB(r_T)}\right]^{2/5} (c/c_s)^{4/5}.
\end{equation}

\cite{HK18} tested this hypothesis by evolving three isothermal disks with temperatures that spanned a factor of 8.  For all of these disks, the results were consistent with $r_T \propto c_s^{-4/5}$.  This scaling held for both disks with MHD turbulence and inviscid HD disks.  Using the measured values of $d\sin \beta/d\ln r$ to calculate $B(r_T)$, as well as the computed value of  ${\cal I} \simeq 0.3$, the observed steady state alignment front locations implied that the dimensionless factor $A$ multiplying the dimensional form $c_s^2/\Omega$ is close to a constant $\simeq 2$.

Interestingly, the form of our relation for the radius of the steady-state transition front may be derived from dimensional analysis \citep{NP00}.  Therefore, it is common to almost all predictions of its location \citep{Pringle92,SF96,II97,NP00}.  Where our model differs is in the dimensionless factor multiplying $c_s^2/\Omega$. All the earlier work supposed that stresses similar to the accretion stress were responsible for regulating these radial mixing motions; consequently, their diffusion rates also involved $\alpha$ ($\propto \alpha^{-1}$ when the warp was small).  However, when phenomenological isotropic viscosity is replaced by physical MHD turbulence, the stresses acting on the radial motions are entirely unrelated to the level of the accretion stress, and the primary influence of MHD turbulence on warp dynamics is through disruption of bending wave propagation, thereby sustaining precession phase gradients \citep{SKH13b}\footnote{It can also play a subtler role in damping oscillations in the location of the steady-state alignment front, as shown by \citet{HK18} and illustrated further in this paper.}  Nonetheless, our model shares with all the earlier ones the same scaling with {\it dimensional} quantities such as sound speed.

Diffusion is not the only process governing disk alignment, however.  Alignment also requires the maintenance of a precession phase gradient against the tendency of angular momentum diffusion to enforce solid-body precession.  In these models, once a steady-state is reached, a precession phase gradient is maintained inside the alignment front, while in the region exterior to the front, the disk maintains its misalignment while precessing as a  solid body; its precession frequency is the Lense--Thirring frequency at the angular momentum-weighted mean radius of the disk (the finite size of our simulated disks influences the outcome in this respect).  Solid body precession can be delayed by the disruption of bending wave propagation, and this is effectively accomplished in MHD models by fluid turbulence stirred by the MRI.  The situation is somewhat different for HD models,  which remain laminar, but bending wave propagation can still be disrupted if disks are cool enough that the disk warp is significantly nonlinear, i.e., $\psi = |\partial \hat\ell /\partial \ln r | > h/r = c_s/v_{\rm orb}$.  The one simulated model in \cite{HK18} that failed to significantly align was the thickest HD model ($h/r \sim 0.1$), for which the $12^\circ$ tilt was insufficient to make the bending waves nonlinear.

The simulations in these previous investigations all used the same tilt angle, $12^\circ$.  How might the alignment process and location of $r_T$ depend on tilt angle?  One direct effect of tilt angle is to determine the magnitude of the warp $\psi$.  As we have found, the ratio of $\psi$ to disk thickness $h/r$ can be significant in determining wave dissipation, and in promoting (or not) the evolution toward solid body precession.   There are also several other ways in which the magnitude of the tilt might influence alignment.  The precession frequency is itself independent of tilt; however, the amplitude of the torque is proportional to it.  Our model for the alignment front velocity depends on $\beta$ only through the integral $\mathcal I$, in which only the ratio between the local $\beta$ and the tilt at the head of the front appears.  Hence we would expect that any tilt dependence of the alignment front velocity should be indirect.  (This property is also shared by the earlier analytic theories, and for essentially the same reason: the magnitude of the Lense-Thirring torque providing the aligning angular momentum is proportional to the magnitude of the tilt that it must correct).  Similarly, the diffusion coefficient $B$ depends on the logarithmic slope of $\beta$ with respect to $r$, which may not have a strong tilt dependence.  Lastly, it is possible that the angle $\gamma$,  which describes the projection of the transported angular momentum onto the aligning direction, might similarly be indirectly affected by tilt angle.

\subsection{New Results}

\begin{figure}[h]
\begin{center}
\includegraphics[width=0.5\textwidth]{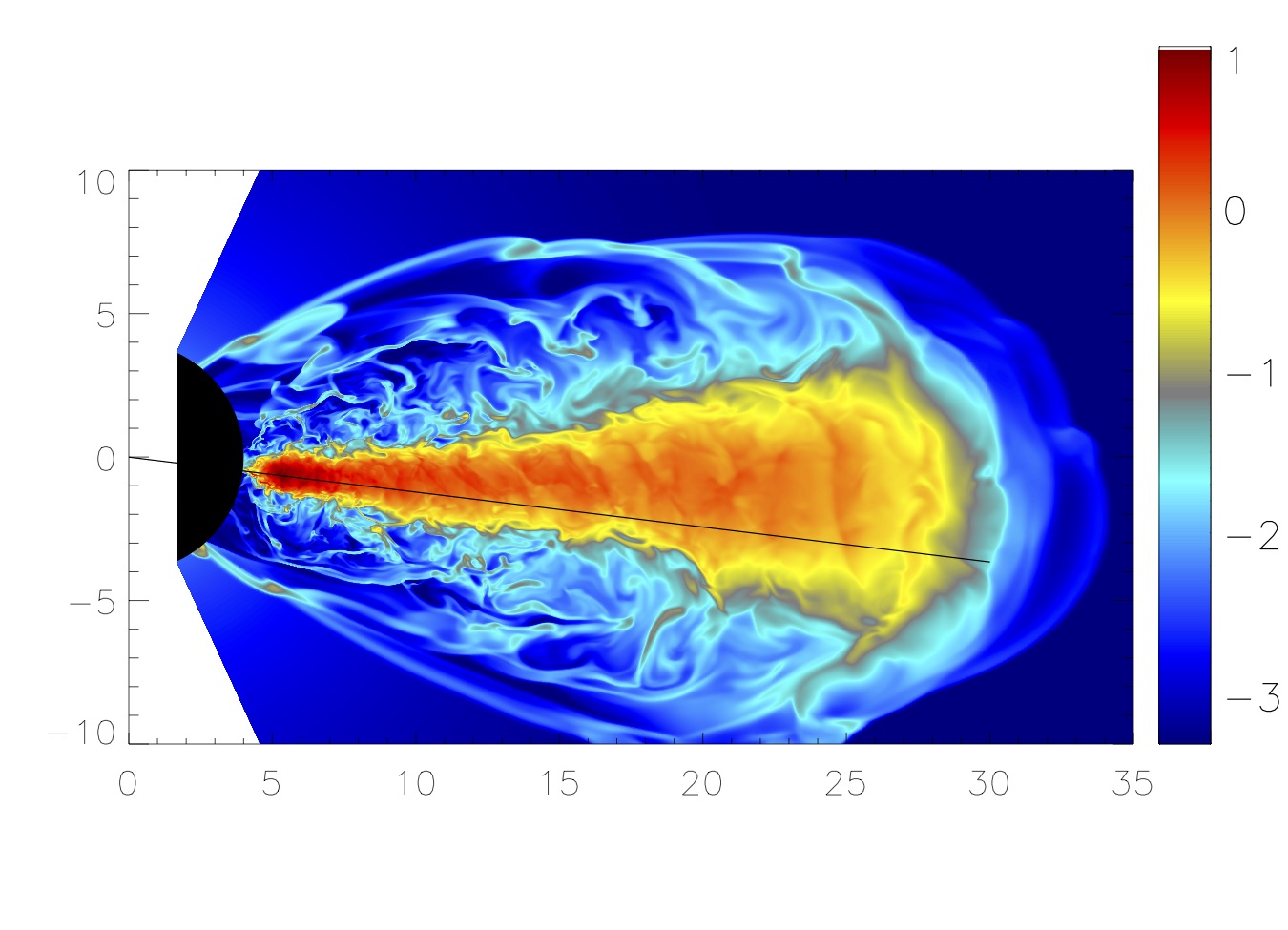}
\includegraphics[width=0.5\textwidth]{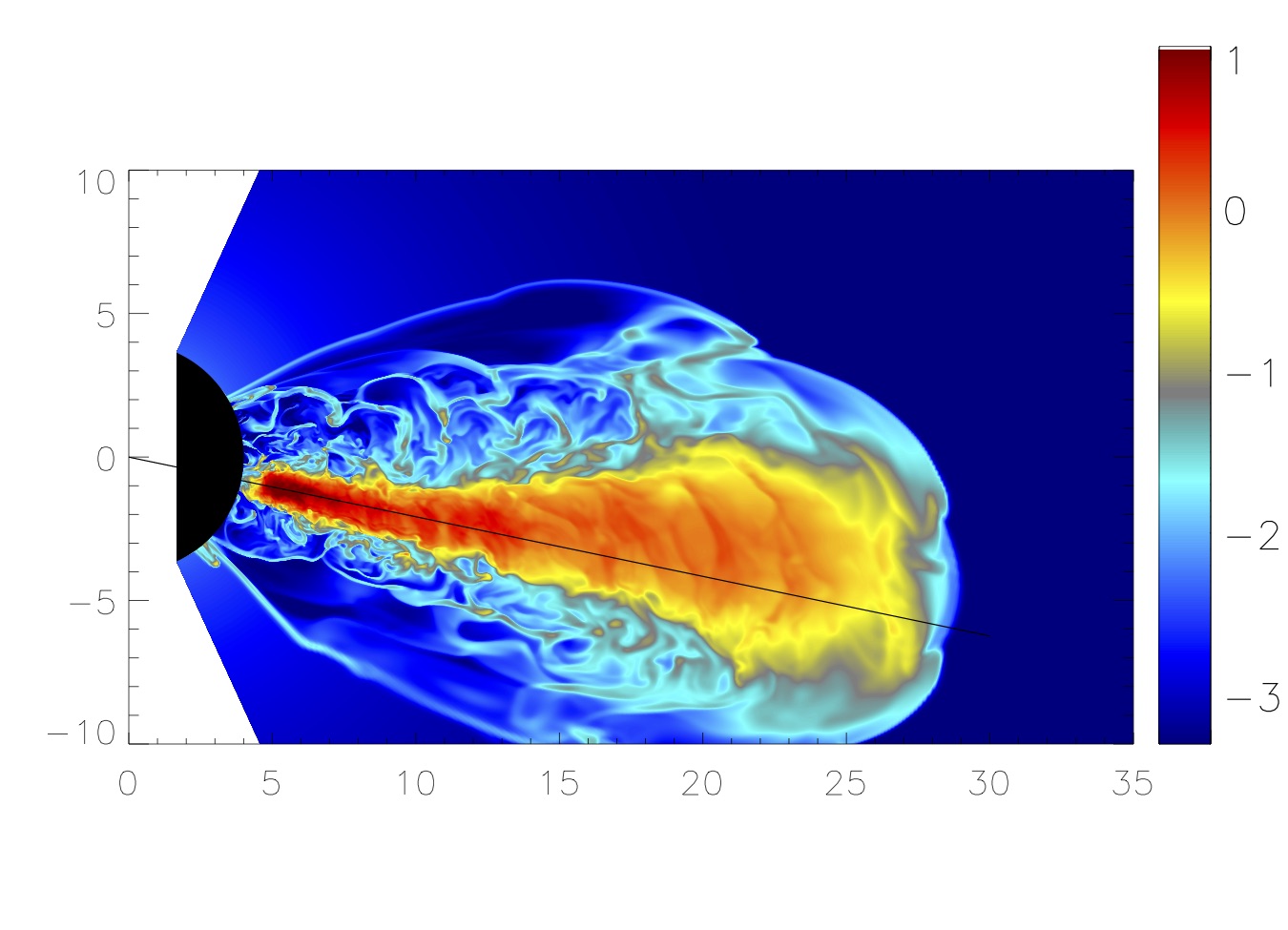}
\includegraphics[width=0.5\textwidth]{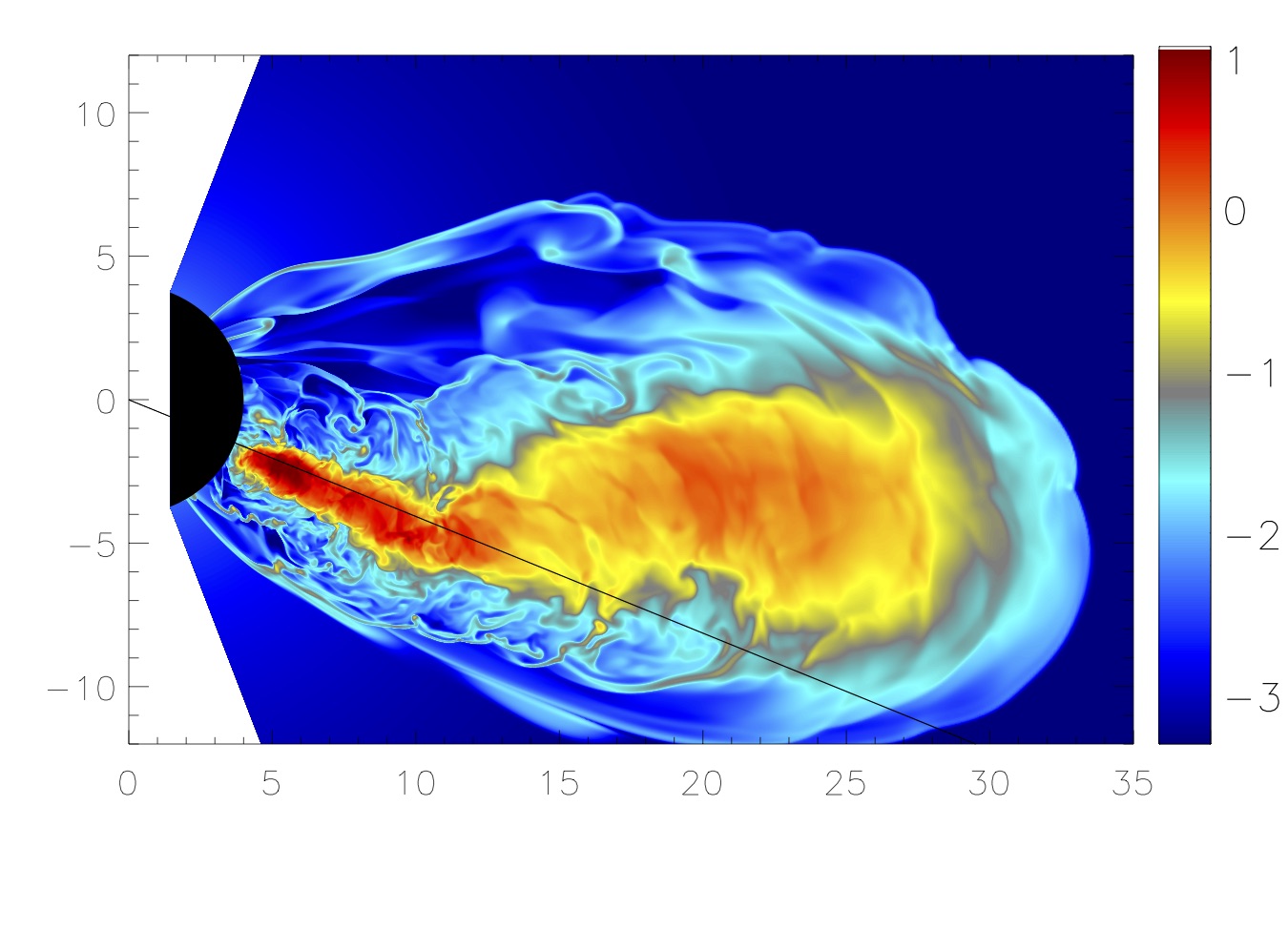}

\caption{Contour plots of log density in the $\phi=0$ plane at orbit 19.5 in Tilt6,  orbit 20 in Tilt12 and orbit 20 in Tilt24.  The range in log density is  from 1.0 to -3.3.   Overlaid is a line showing the equatorial plane for the spin axis of 6, 12 and 24 degrees, as appropriate.   }
\label{fig:tilt24-density}
\end{center}
\end{figure}

Figure~\ref{fig:tilt24-density} shows a density contour slice for the three MHD models with  tilt angles $6^\circ$, $12^\circ$ and $24^\circ$ in the $\phi =0$, i.e., $x$-$z$, plane after ~20 orbits of torque.  At a qualitative level, it is apparent that the inner regions of all three models have aligned with the Lense--Thirring equatorial plane, and the transition from inner alignment to outer obliquity takes place at approximately the same radius.
   
\begin{figure}[h]
\begin{center}
\includegraphics[width=0.5\textwidth]{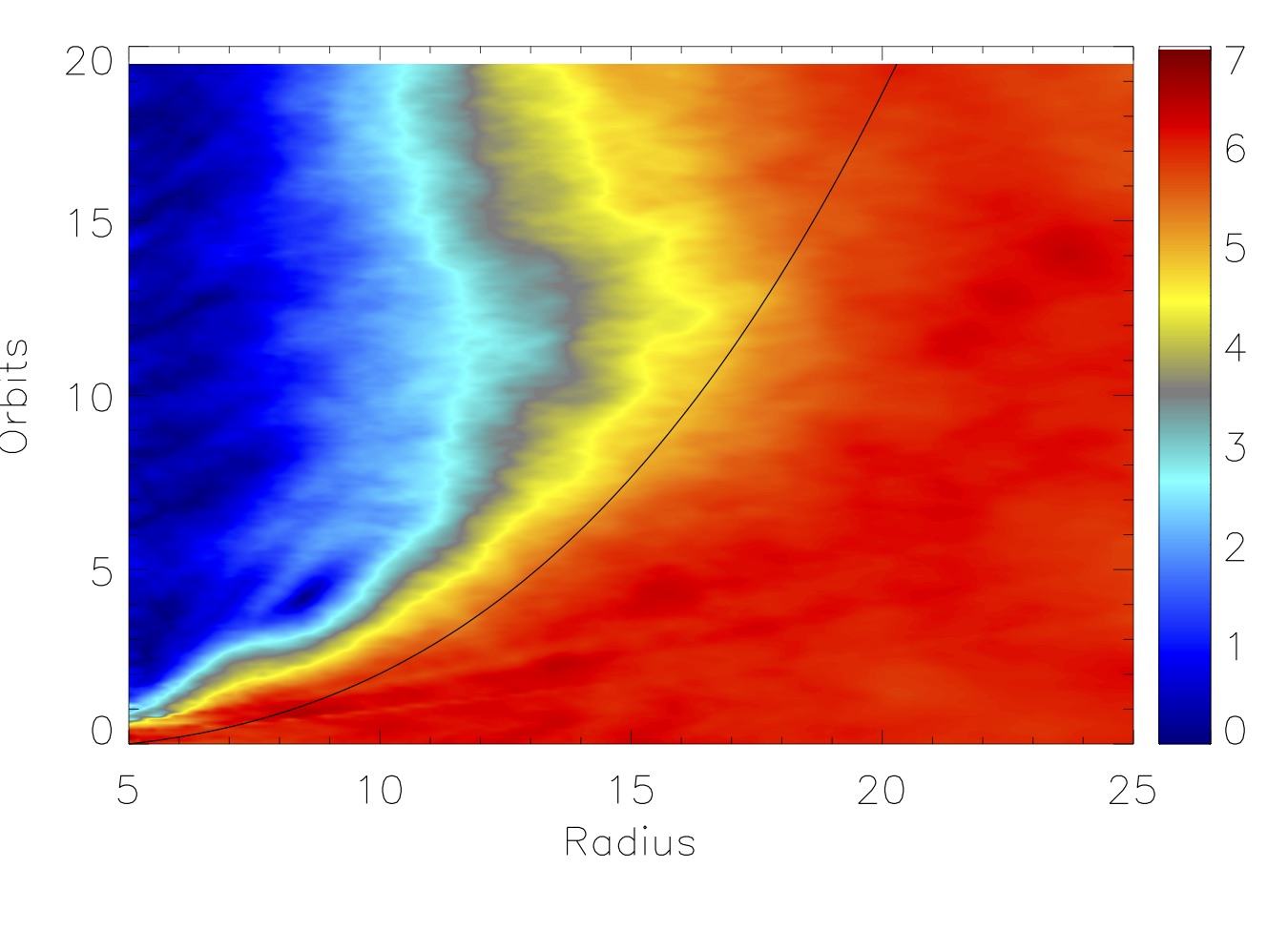}
\includegraphics[width=0.5\textwidth]{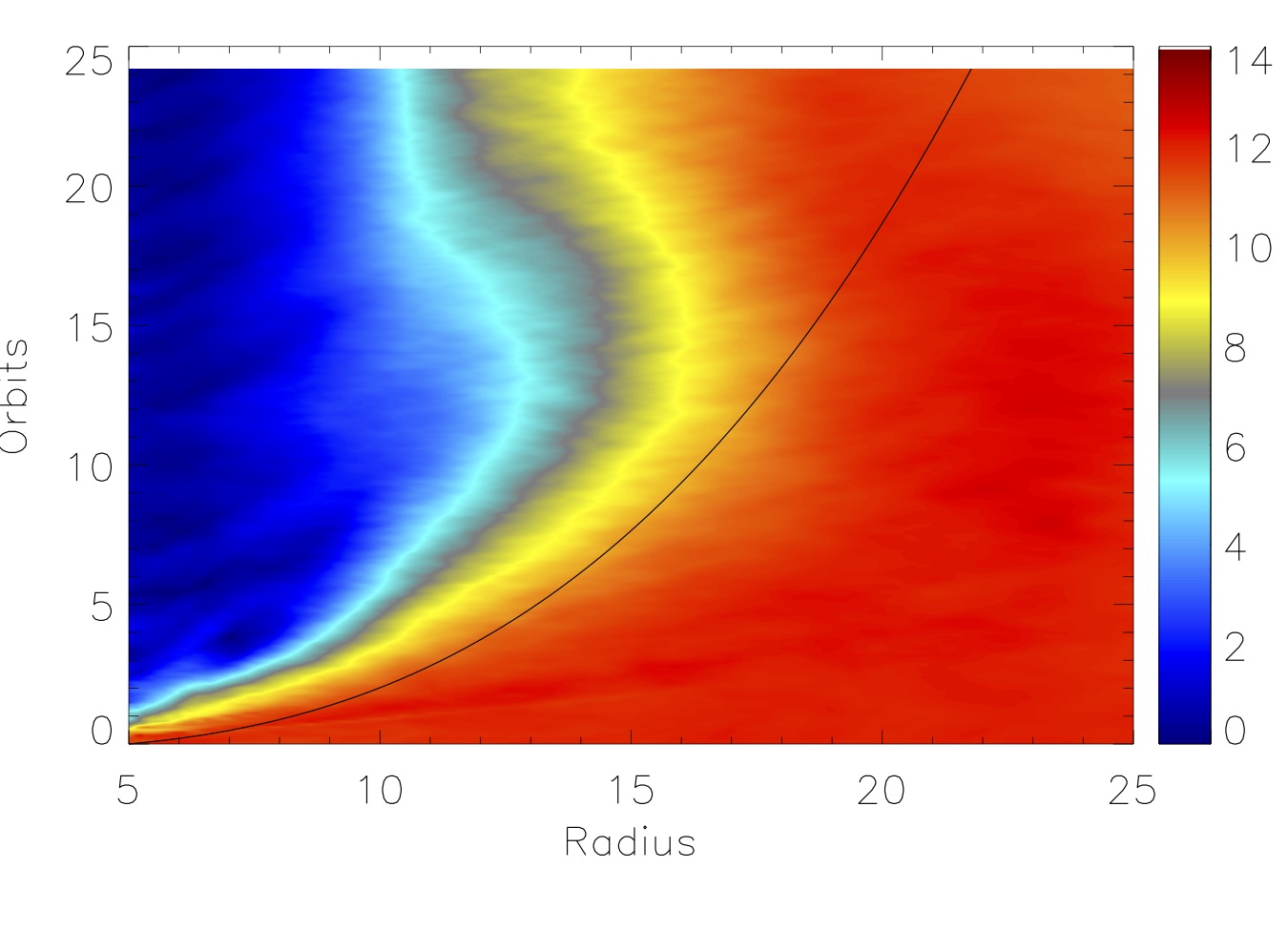}
\includegraphics[width=0.5\textwidth]{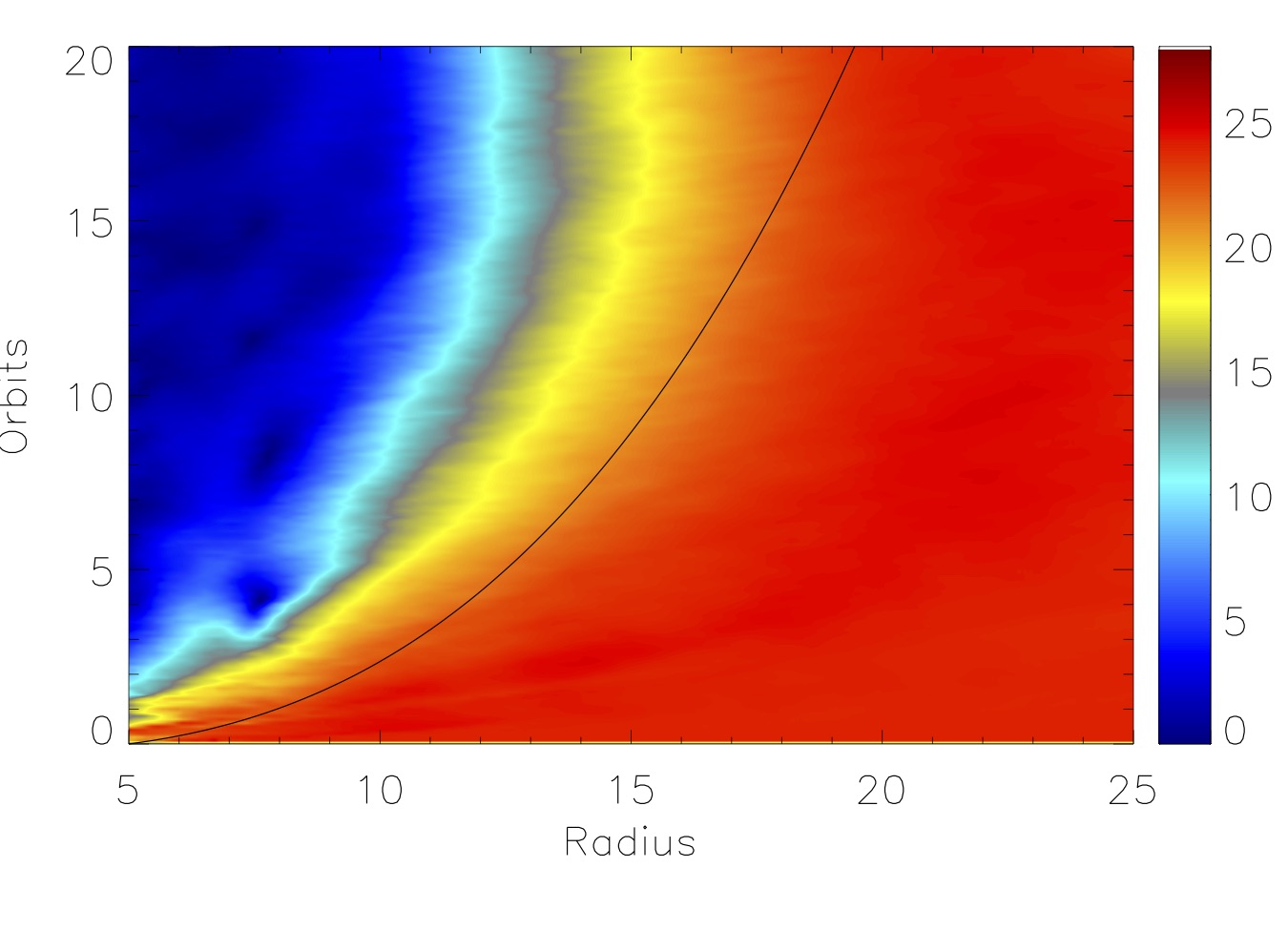}
\caption{Spacetime diagram for the alignment angle $\beta$ in the Tilt6 (top), Tilt12 (middle) and Tilt24 (bottom) model.   The color scale runs from $\beta=0$ (blue: aligned) to $\beta=7$, 14 and 28 degrees (red) respectively so that the same color corresponds to the same percentage of alignment from model to model.  Overlaid on the spacetime diagrams is a curve corresponding to an alignment front speed of $0.35 r\Omega_{\rm precess}$ for Tilt6 and Tilt24, and $0.3 r\Omega_{\rm precess}$ for Tilt24.
}
\label{fig:tilt24beta}
\end{center}
\end{figure}

Figure~\ref{fig:tilt24beta} presents the space-time evolution of the alignment angle $\beta$ in each of the three MHD models.  The color scales are proportional to the tilt angle so that the colors correspond to the same degree of alignment from model to model; for example, the point where the disk is 50\% aligned corresponds to the cyan/gray boundary in color.  
Vindicating the visual impression given in Fig.~\ref{fig:tilt24beta}, the half-alignment radius varies  somewhat with tilt angle, but not in a systematic way: $r_T \approx 14$ for Tilt6, $r_T \approx 11.5$ for Tilt12, $r_T \approx 13.5$ for Tilt24.   Moreover, there is also considerable similarity in the alignment front's initial evolution.   In each case, the alignment front travels outward at a velocity $\propto r \Omega_{\rm precess}$.    If the front position is defined by 20\% alignment,  the alignment front's speed in these units is $\approx  0.35$ in Tilt6 and Tilt12 and 0.30 in Tilt24. The time at which the initial alignment front stalls is nearly the same for Tilt6 and Tilt12, but, consistent with a slower alignment front velocity, stalling  occurs about 5 orbits later in Tilt24.

\begin{figure}[h]
\begin{center}
\includegraphics[width=0.7\textwidth]{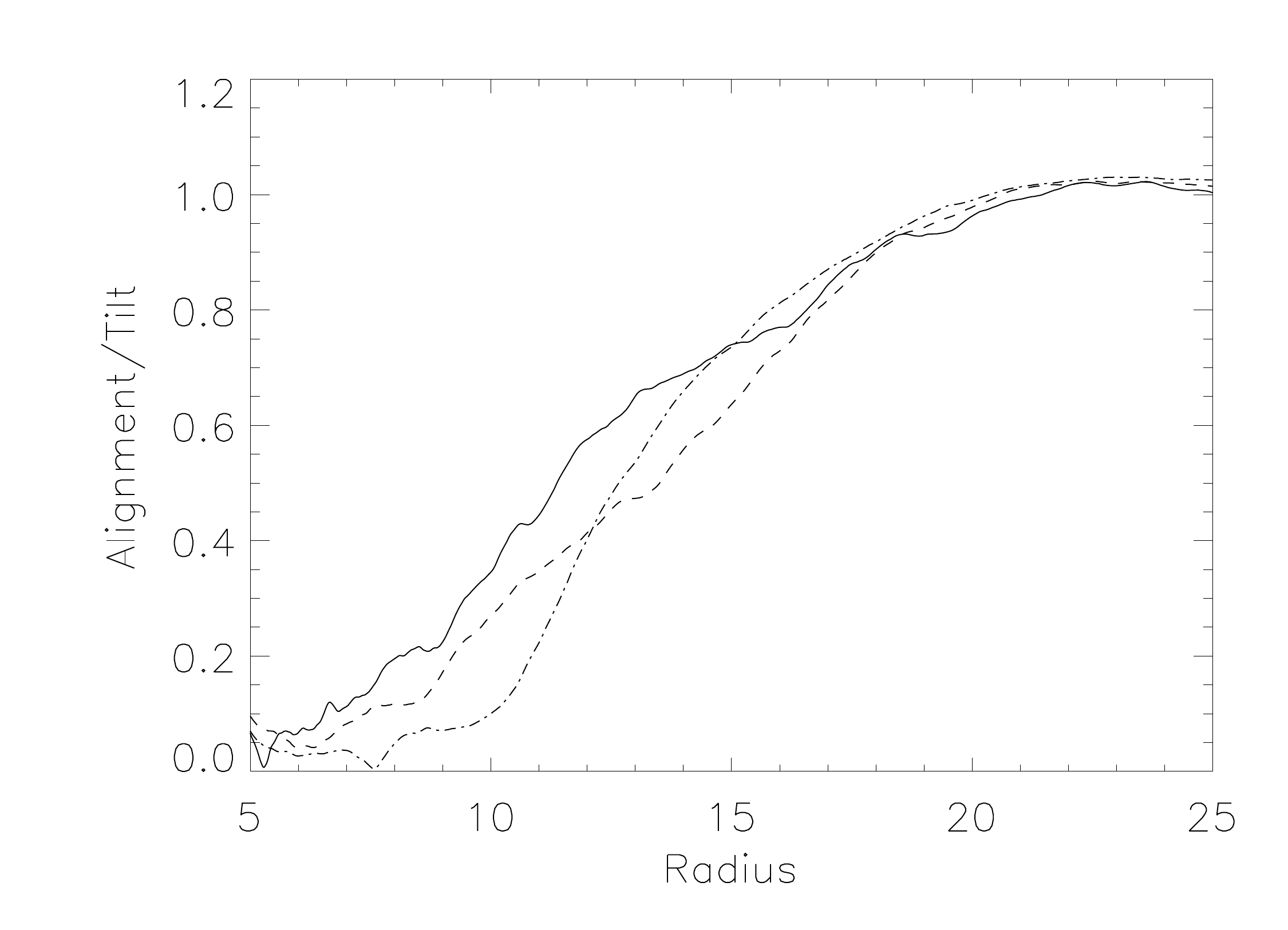}
\caption{Plot of the radial dependence of the alignment angle $\beta$ divided by the tilt angle $\beta_0$ at orbit 15 for Tilt24 (dot-dashed line), Tilt12 (dashed line) and Tilt6 (solid line).   The three tilts show comparable alignment when scaled for the original tilt angle.  There is good alignment inside of $r=10$; beyond this point $\beta$ rises to the original tilt angle.  The outer disk beyond $r=21$ remains unaligned.
}
\label{fig:tilt24beta-comp}
\end{center}
\end{figure}

Figure~\ref{fig:tilt24beta-comp} demonstrates this strong similarity between the three models by showing normalized alignment angle $\beta (r)/\beta_0$ after 15 orbits of torque, where the outer tilt is $\beta_0$. All three tilt angles show near alignment inside of $r=10$, followed by a transition region spanning $r\sim 12$--$22$ as $\beta/\beta_0$ increases smoothly to 1.  The slope of $\partial \sin\beta / \partial \ln r$ between $r=10$ and 20 is  0.10, 0.23, and 0.55 in order of increasing tilt angle.  The slope increases by a factor of $\approx 2.3$--2.4 for each doubling of tilt angle, so the increase is greater than that attributable to the change in tilt alone.   This is visible in the plot: higher tilt values give more complete alignment within the inner disk, resulting in a steeper rise in $\beta$ to the outer, unaligned disk.

\begin{figure}
\begin{center}
\includegraphics[width=0.5\textwidth]{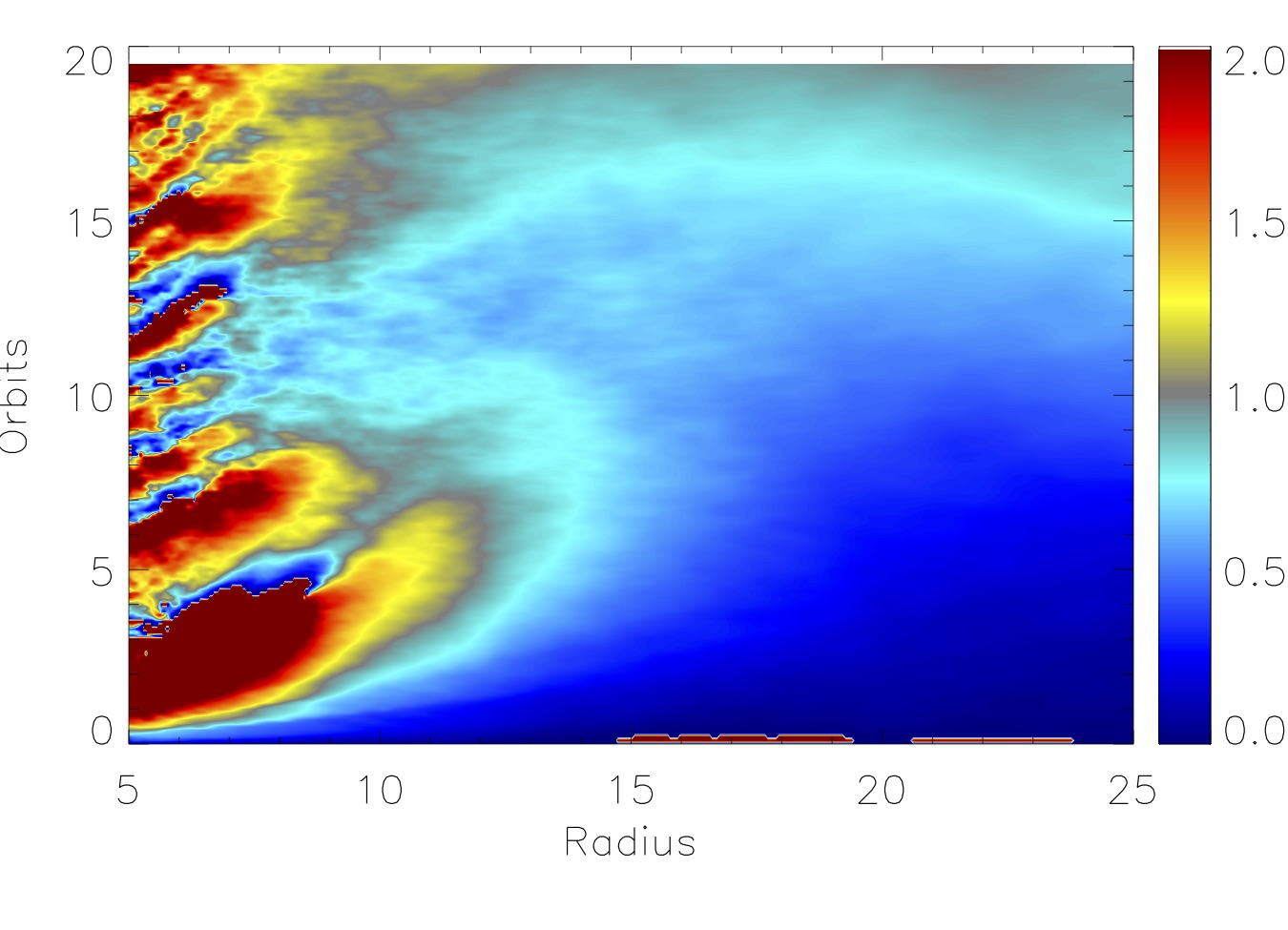}
\includegraphics[width=0.5\textwidth]{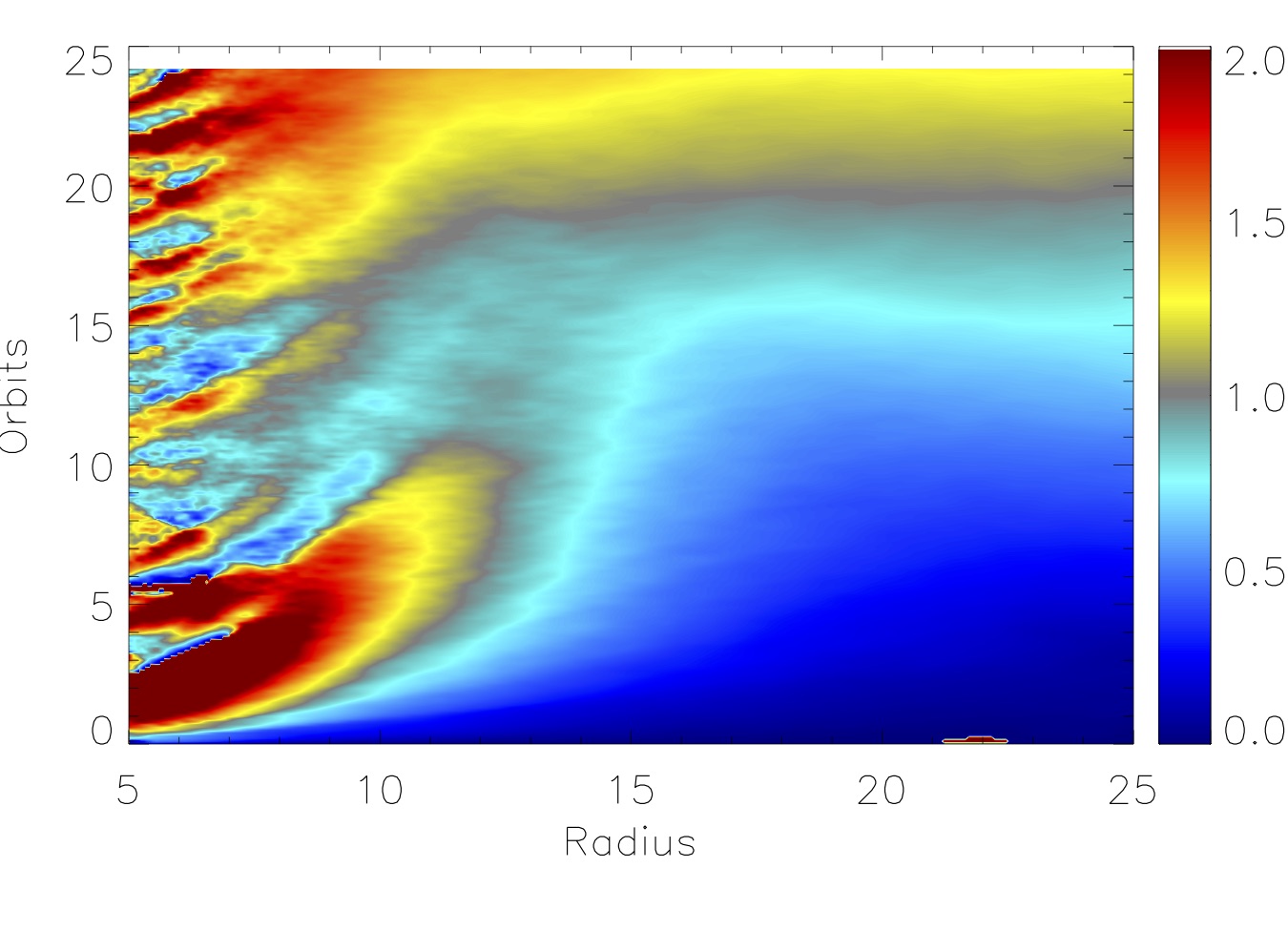}
\includegraphics[width=0.5\textwidth]{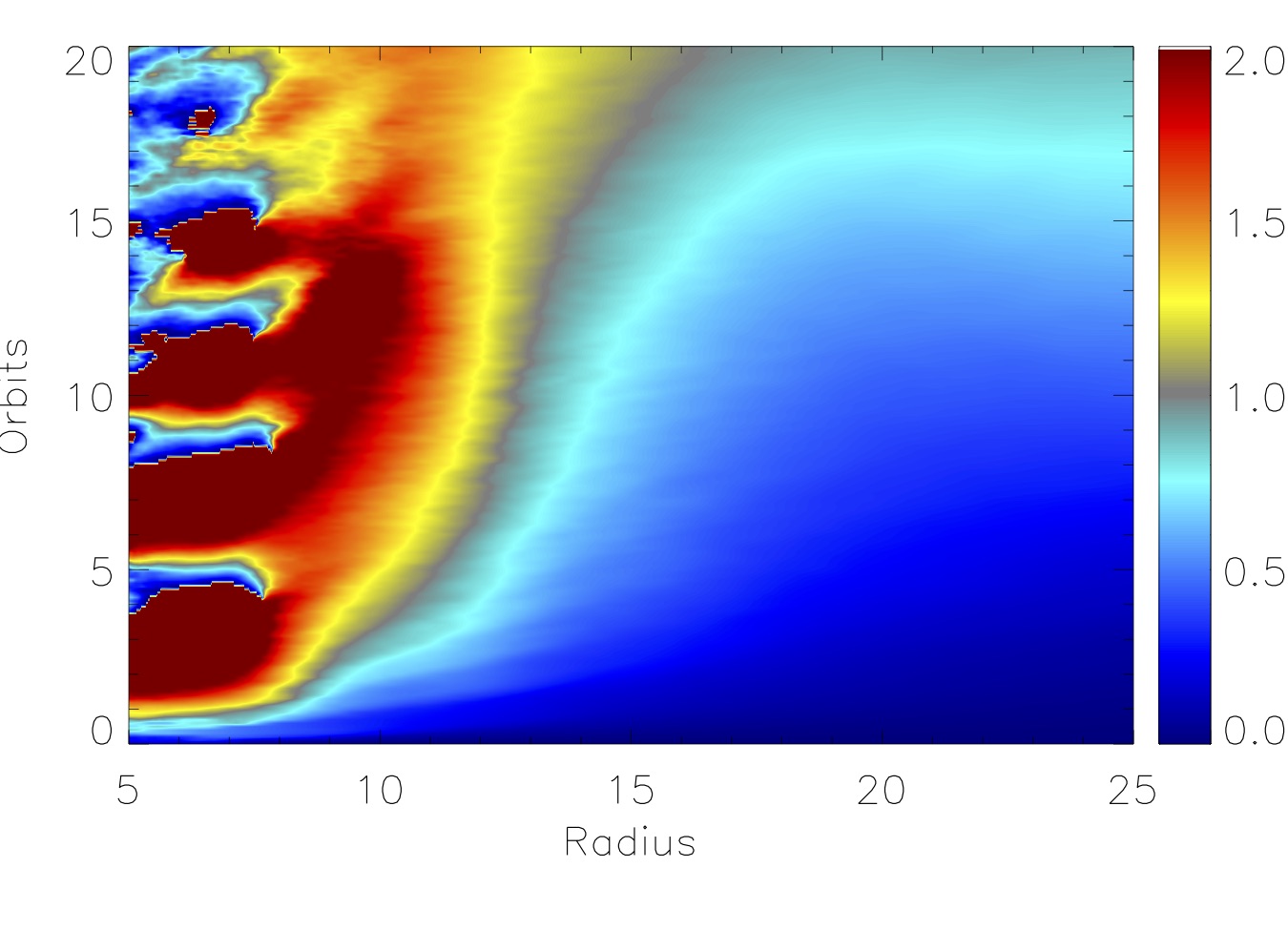}

\caption{Spacetime diagram for the precession angle $\phi$ in the Tilt6 (top), Tilt12 (middle) and Tilt24(bottom) model.  Colors run from $\phi=0$ (blue) to $\phi=2$ radians (red); values between 2 and $2\pi$ are dark red.  
}
\label{fig:tilt24phi}
\end{center}
\end{figure}

Figure \ref{fig:tilt24phi} displays the spacetime diagrams for the precession angle $\phi$.  For the first 5 orbits the precession rate in all three models is the local Lense--Thirring precession rate.   After the first 5 orbits, differences develop between $\Omega_{\rm precess}$ and the observed precession rate.   As the evolution proceeds, the phase angle gradient in the outer disk ($r > 15$) tends toward zero.  This happens at later times for greater tilts, although the differences are not substantial.

\begin{figure}
\begin{center}
\includegraphics[width=0.5\textwidth]{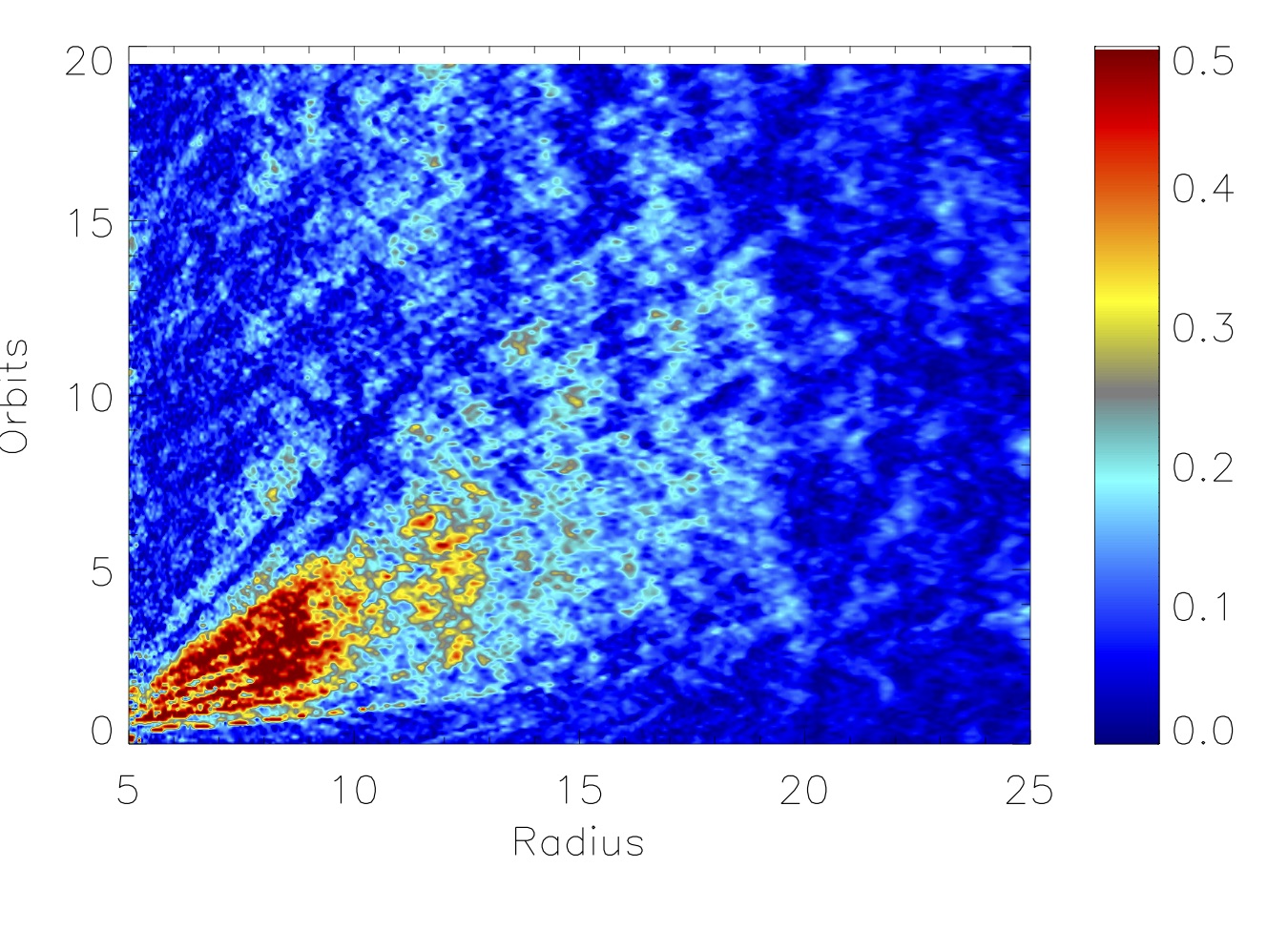}
\includegraphics[width=0.5\textwidth]{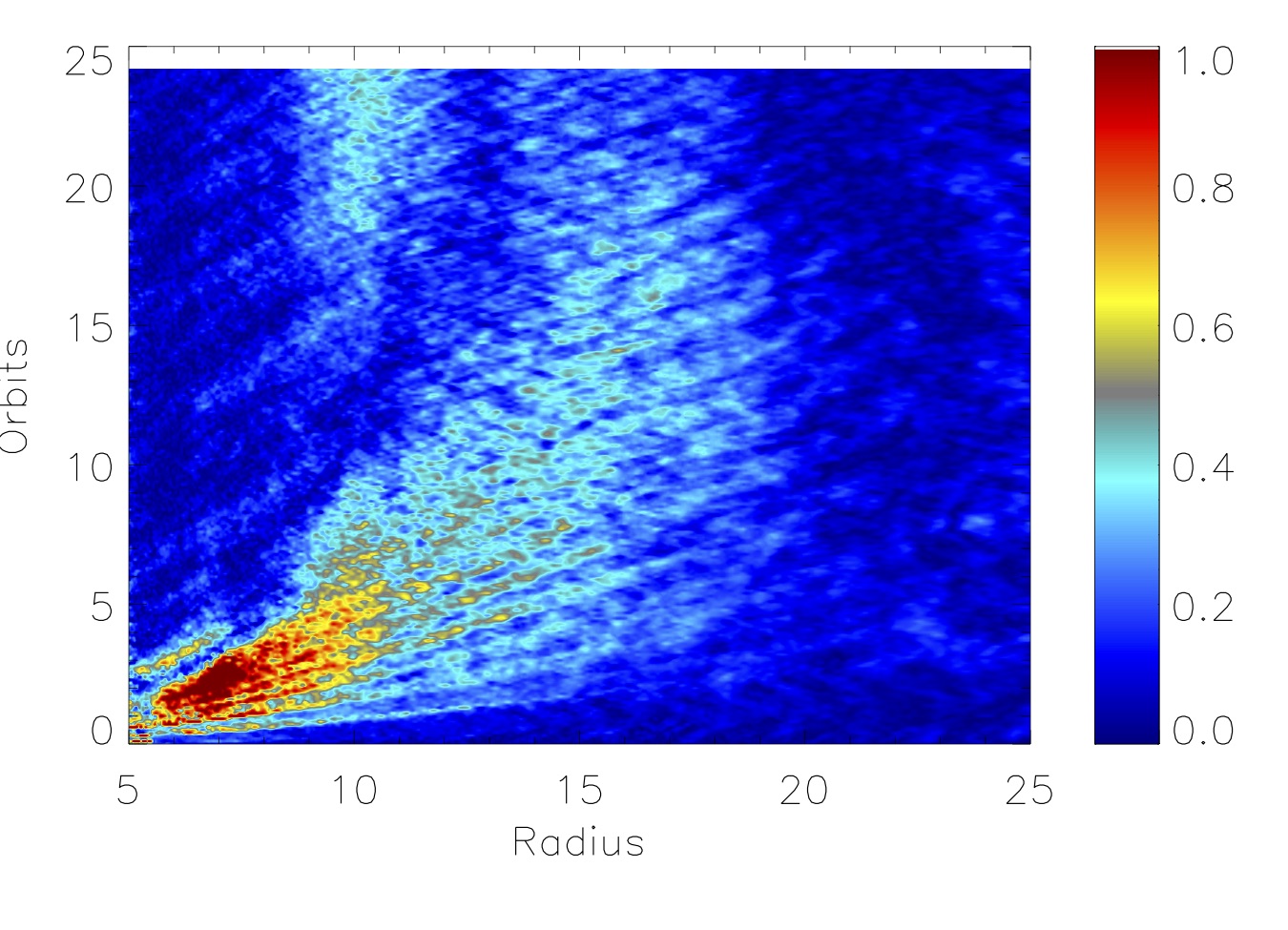}
\includegraphics[width=0.5\textwidth]{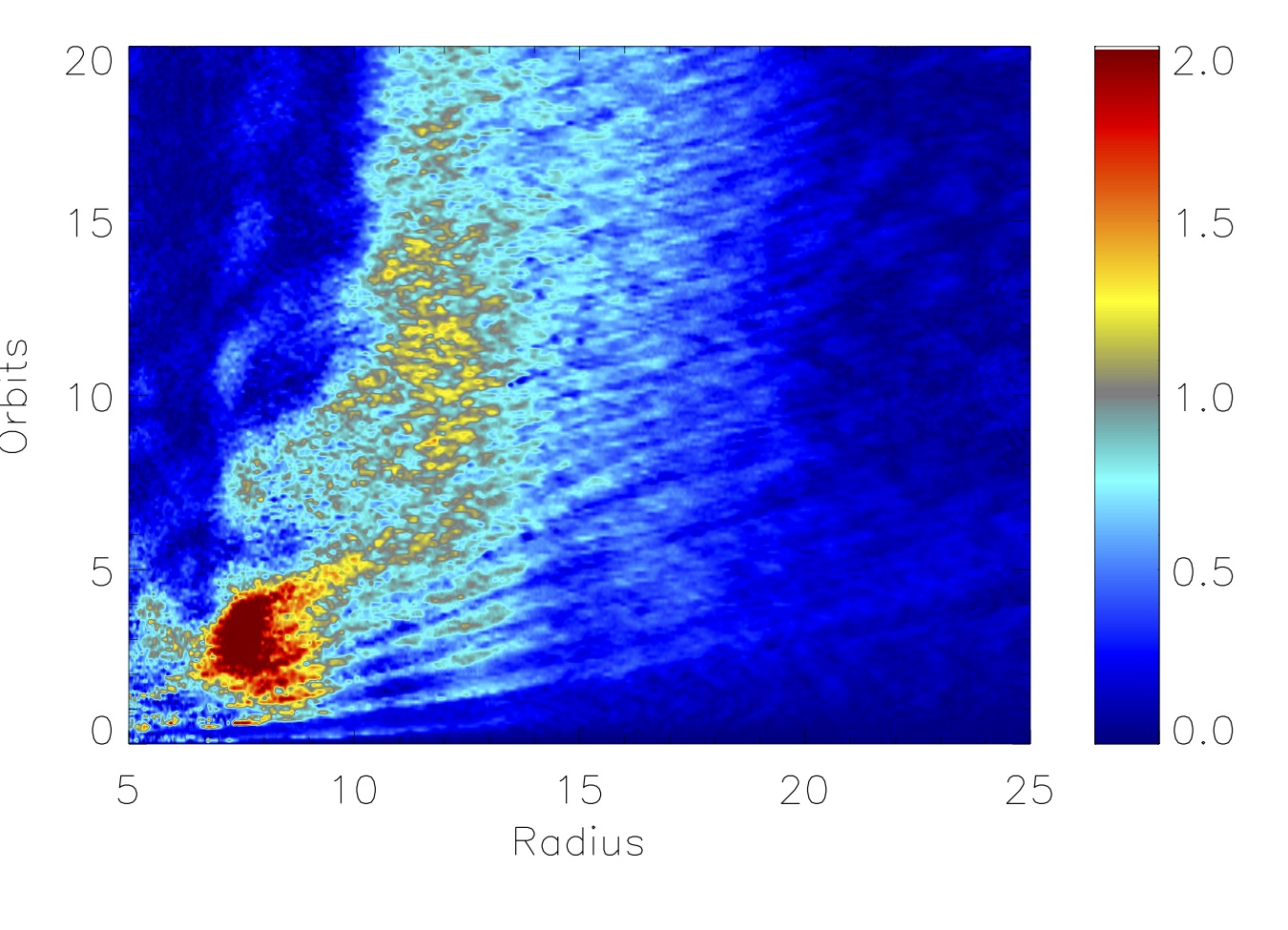}
\caption{Spacetime diagram for the warp $\psi$ in the Tilt6 (top), Tilt12 (middle) and Tilt24 (bottom) model.  Colors run from $\psi=0$ (blue) to a maximum (red) of $\psi=0.5$ for Tilt6, 1.0 for Tilt12 and 2.0 for Tilt24.
}

\label{fig:tilt24psi}
\end{center}
\end{figure}

Figure \ref{fig:tilt24psi} is the spacetime diagram of warp $\psi$ in the three MHD runs.  Since all three models have the same sound speed, the conversion of $\psi$ into normalized warp $\hat\psi = \psi/(h/r)$ is essentially the same.  Note that the color scale in each figure increases proportional to the increase of the tilt angle.  Even with this increase in scale, however, it is evident that $\psi$ increases with tilt beyond what is attributable to the increase in tilt alone.  For each model there is an initial period during the first 5 orbits of strong nonlinear warp amplitude inside $r=10$.   The relatively large amplitude of $\psi$ in this initial phase arises with the impulsive start of the torque at the beginning of the simulation. Even for the $6^\circ$ tilt, the amplitude of the initial imposed warp is large compared to the disk thickness at $r=5$, which at  $h/r = 0.035$ gives a disk opening angle at that radius of $\sim 4^\circ$.  These initial waves move outward through the disk, losing strength as they do so.  After the departure of these initial waves, Tilt6 has a low value of $\psi$, consistent with the slope of $\beta$.   Tilt24 has a relatively larger $\psi$ in the alignment region outside of $r=10$, consistent with the steeper slope of the alignment angle $\beta$.  The behavior of Tilt12 is intermediate between the other two models. In all three cases after the first five orbits, the outward propagating coherent bending waves are only marginally evident and at a low amplitude.  

\begin{figure}
\begin{center}
\includegraphics[width=0.7\textwidth]{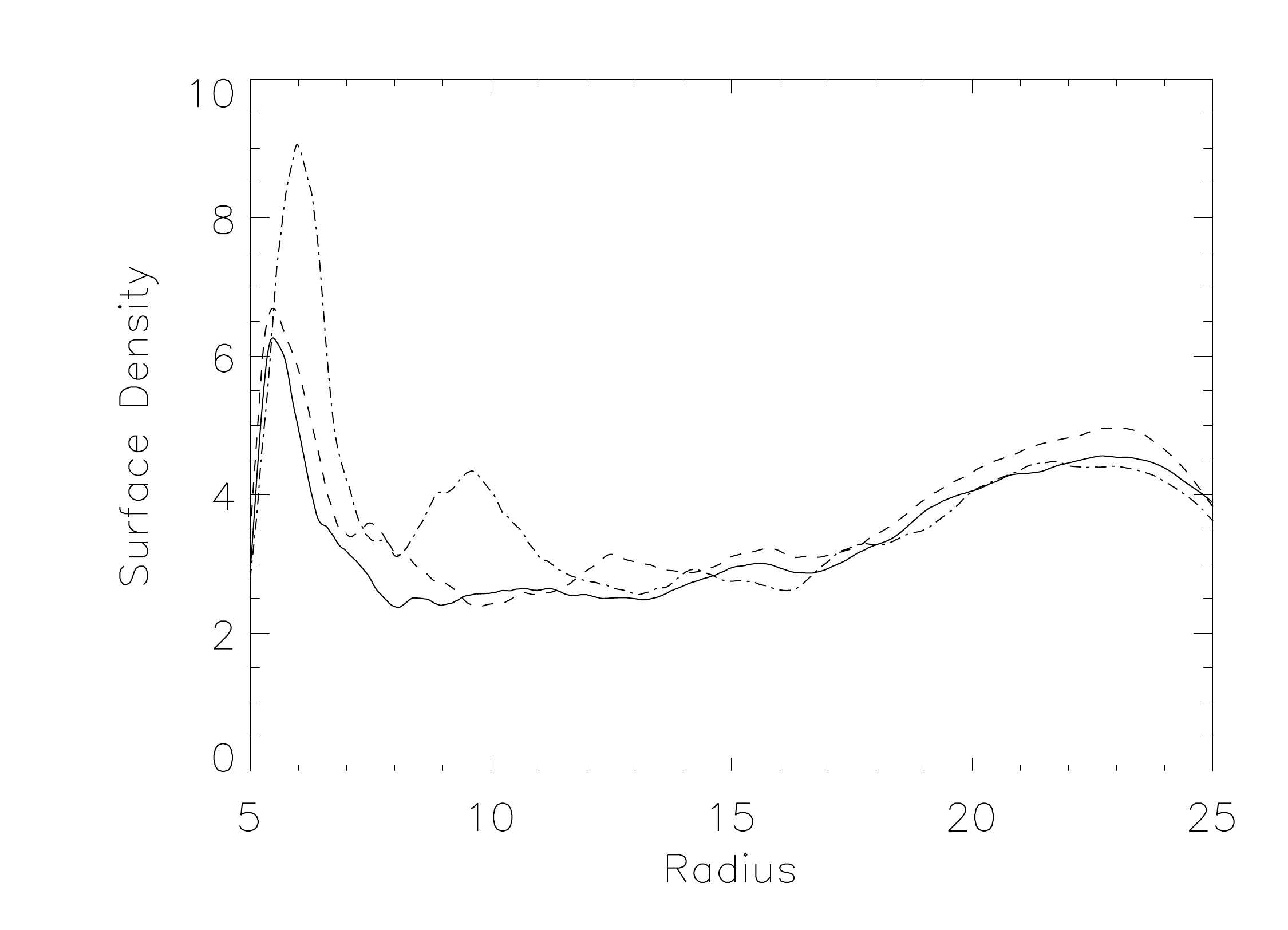}
\caption{Plot of the surface density $\Sigma$ as a function of radius for Tilt6 (solid line) at 19.5 orbits, Tilt12 (dashed line) at 20 orbits, and Tilt24 (dot-dashed line) at 20 orbits. In all three models $\Sigma$ is smooth and gradually increasing from $r=12$ to $r \sim 24$, the radial range over which the disks transition from aligned to the original disk tilt.  There is no evidence of disk breaking.
}
\label{fig:sigma}
\end{center}
\end{figure}

As mentioned in Sec.~1, there has been considerable discussion of whether disks misaligned by a large angle can ``break", i.e., the outer portion of the misaligned disk becomes physically separated from a (more) aligned inner disk.  Although this phenomenon has been observed in some SPH tilted disk simulations, the reasons for it remain uncertain.  We see no evidence for disk breaking, or even the initial onset of breaking, in our finite-difference simulations,  even when $\beta_0 = 24^\circ$.  Figure \ref{fig:tilt24-density}, a density slice of the three tilt models at 20 orbits after the onset of torque, exhibits no systematic reduction in density through the alignment region that lies between the aligned inner disk and the unaligned outer disk.  Figure~\ref{fig:tilt24beta-comp} shows that $d\beta/dr$ remains smooth and continuous, even as the slope steepens with increasing tilt.  Figure~\ref{fig:sigma} shows the azimuthally averaged disk surface density $\Sigma$, computed on spherical shells as a function of radius for the three models at $\sim 20$ orbits.  A few modest differences can be seen from one tilt to the next in the aligned inner disk, but the three models are very similar to one another in the transition region, which runs from $r=12$ to 24.  $\Sigma(r)$ is smooth and shows no evidence for a systematic reduction in value at any radius within the transition region.  

In summary, we see that smooth alignment is achieved in very similar fashion for tilt angles ranging from $6^\circ$ to $24^\circ$.   There are some  minor quantitative differences between the models.  The slope  $\partial \sin\beta / \partial \ln r$  increases with tilt by more than any increase due solely to the imposed tilt. That is, the transition from the inner aligned disk to the outer unaligned disk becomes relatively steeper with greater tilts. The outward traveling bending waves produced by the initial impulsive onset of the torque in all three models are large relative to $h/r$, but after that initial phase, the amplitudes of outward traveling bending waves are small.  We see no evidence for disk breaks, or even incipient breaks, across this range of tilt angles, despite the thin ($h/r=0.05$ at $r=10$) profile of the disk.   

\subsection{Hydrodynamic models}

\begin{figure}
\begin{center}
\includegraphics[width=0.5\textwidth]{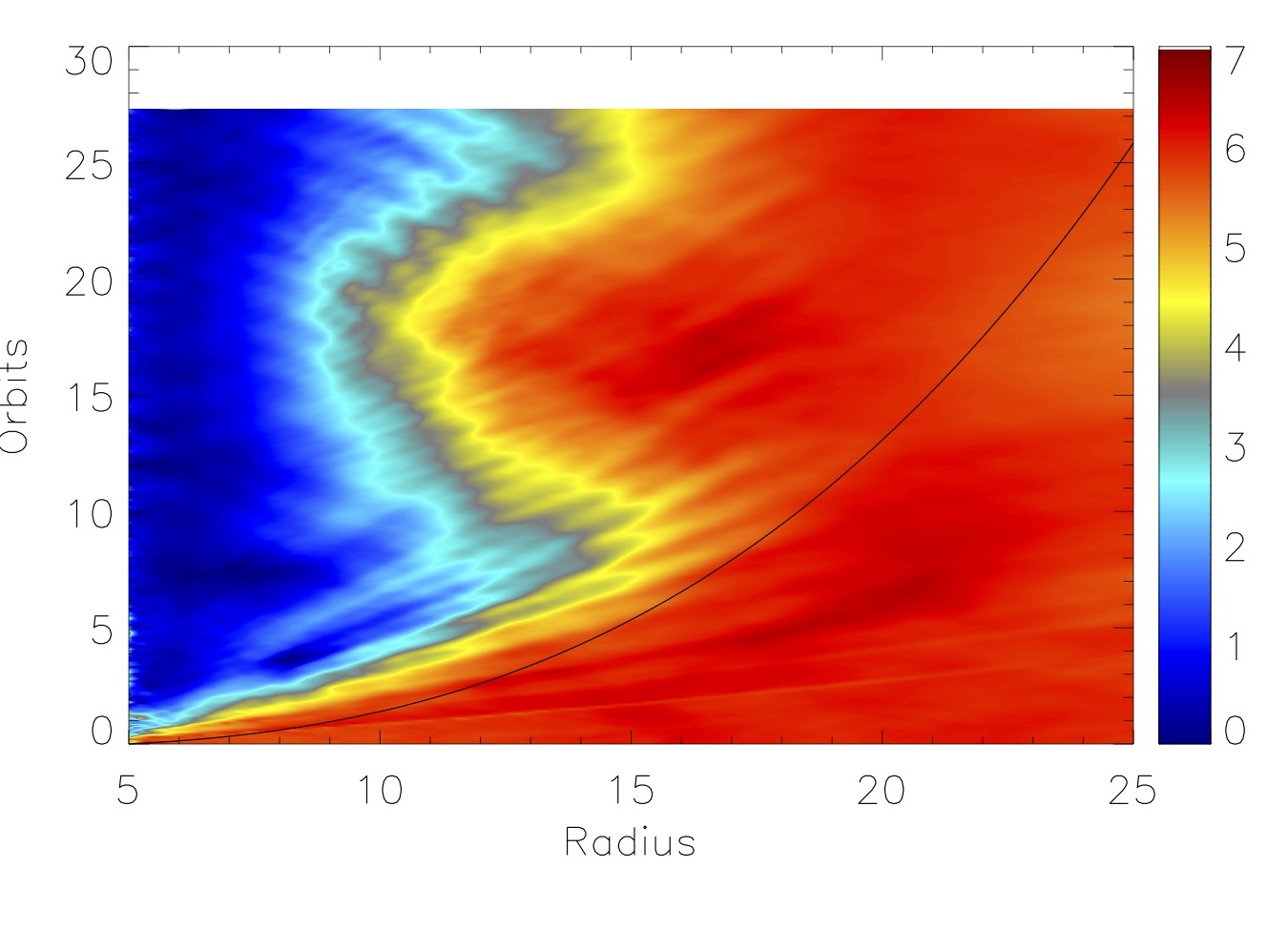}
\includegraphics[width=0.5\textwidth]{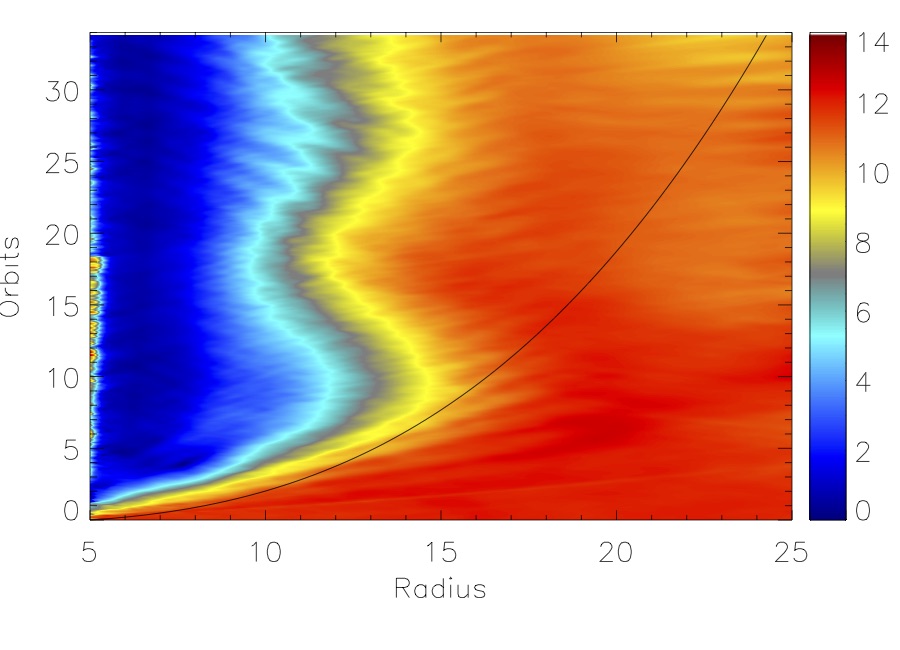}
\includegraphics[width=0.5\textwidth]{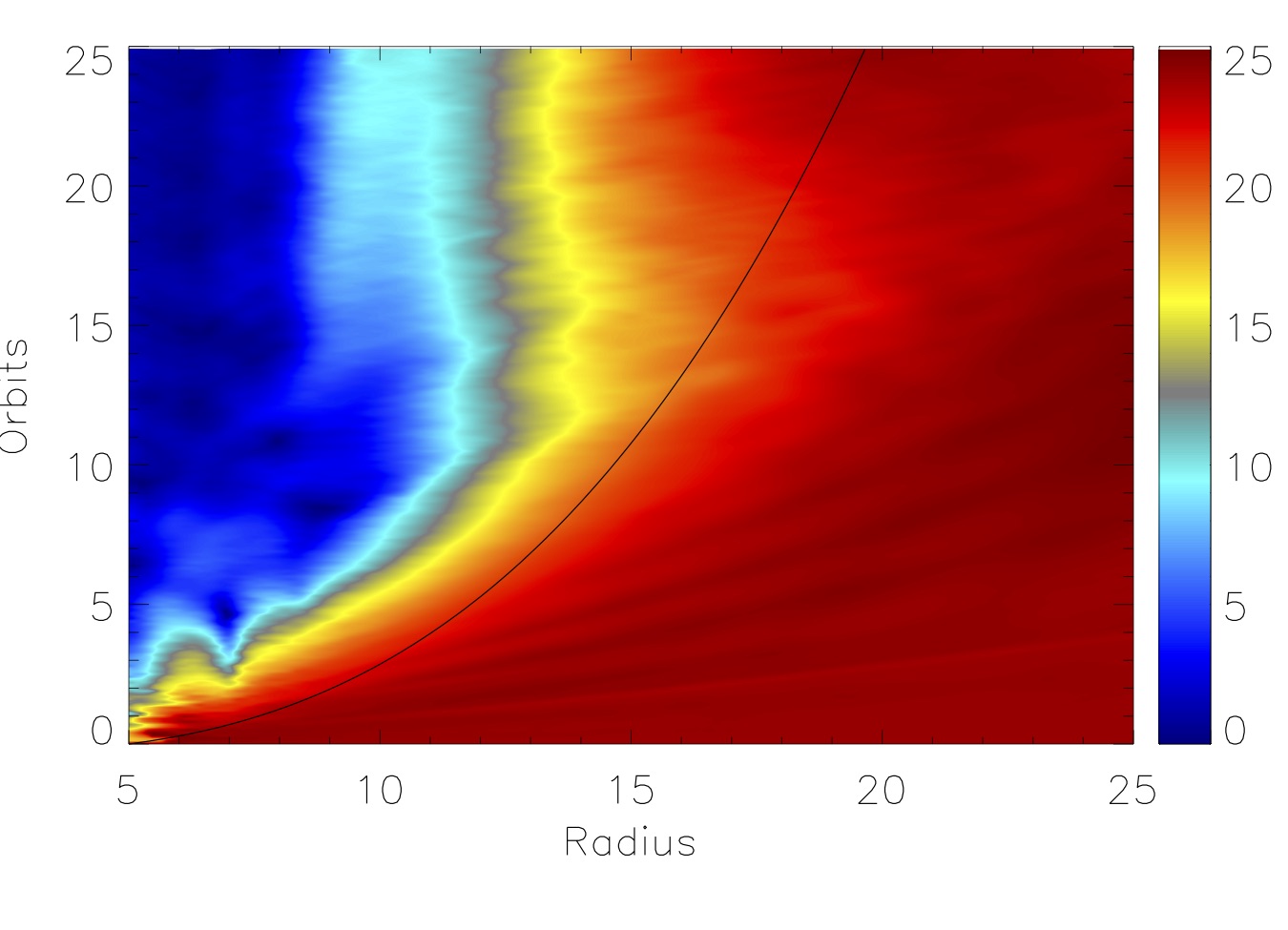}
\caption{Spacetime diagram for the alignment $\beta$ for the hydrodynamic models Tilt6-H (top), Tilt12-H (middle) and Tilt24-H (bottom).  In each plot a line shows a trajectory moving through spacetime with a velocity of $0.35 r\Omega_{\rm precess}$ (top, middle) and $0.25 r\Omega_{\rm precess}$ (bottom). }
\label{fig:betaH}
\end{center}
\end{figure}

We next turn to the hydrodynamic counterparts to the three MHD models.
The spacetime diagrams for alignment $\beta$ are given in Figure \ref{fig:betaH}.  Qualitatively, they resemble those of the MHD models (Fig.~\ref{fig:tilt24beta}), but several contrasts stand out.  First, significant coherent bending waves (the diagonal linear features in $\beta$) are clearly present in all three cases.   However, although both inward- and outward-traveling waves are seen in Tilt6-H and Tilt12-H, only outward-traveling waves are evident in Tilt24-H.  In addition, the relative amplitude of the waves (after normalizing for tilt) is larger in the smaller tilt cases.  Second, unlike the MHD models, in which the alignment front monotonically approaches its steady-state location after a single excursion, the front's location in Tilt6-H and Tilt12-H oscillates with a period of $\sim 20$ orbits.  The amplitude of the oscillation is larger for the $6^\circ$ case than for the $12^\circ$; in fact, although the oscillation amplitude in the $12^\circ$ case appears to be diminishing over time, this decay is noticeably slower for the smaller tilt. 

\begin{figure}[h]
\begin{center}
\includegraphics[width=0.7\textwidth]{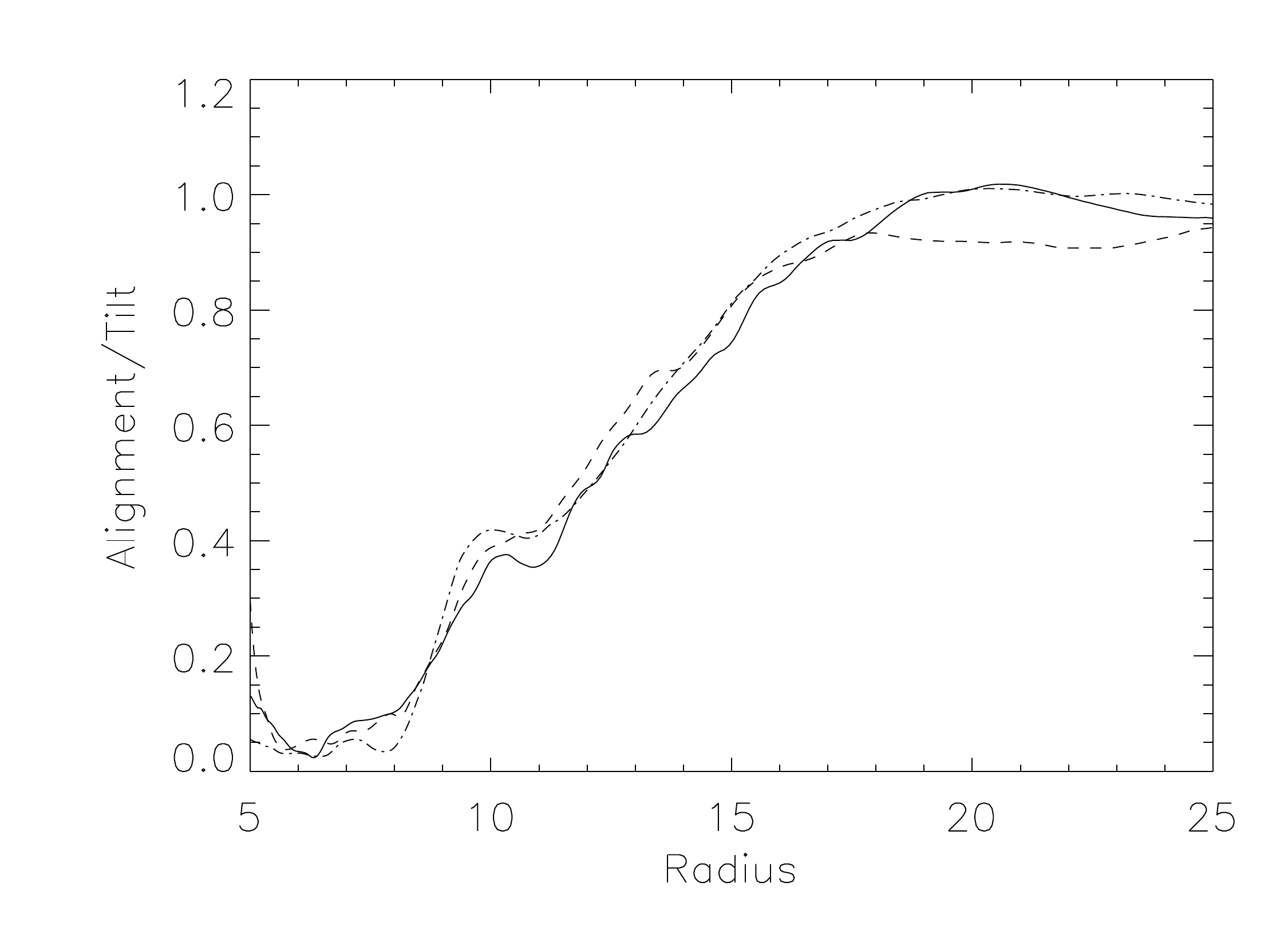}
\caption{Plot of the radial dependence of the alignment angle $\beta$ divided by the tilt angle $\beta_0$ at  27 orbits for Tilt6-H (solid line) and Tilt12-H (dashed line), and 25 orbits for Tilt24-H (dot-dashed line).   The HD models exhibit essentially the same alignment slope when scaled by the tilt angle.  
}
\label{fig:beta-comp-H}
\end{center}
\end{figure}

The slope $\partial \beta/\partial \ln r$ between $r=7$ and 20 is 0.10, 0.22 and 0.41 for the three tilt angles, i.e., increasing linearly with tilt.   Figure~\ref{fig:beta-comp-H} demonstrates that the slope is essentially linear in $\beta_0$ at the end of each of the evolution for these models.  There is even a similarity in the locally flat feature that can be seen near $r=10$.  The value of $r_T$, defined as the point where $\beta$ equals half the tilt angle, is $r_T = 12.1$, 11.6 and 12.3 respectively.  

\begin{figure}
\begin{center}
\includegraphics[width=0.5\textwidth]{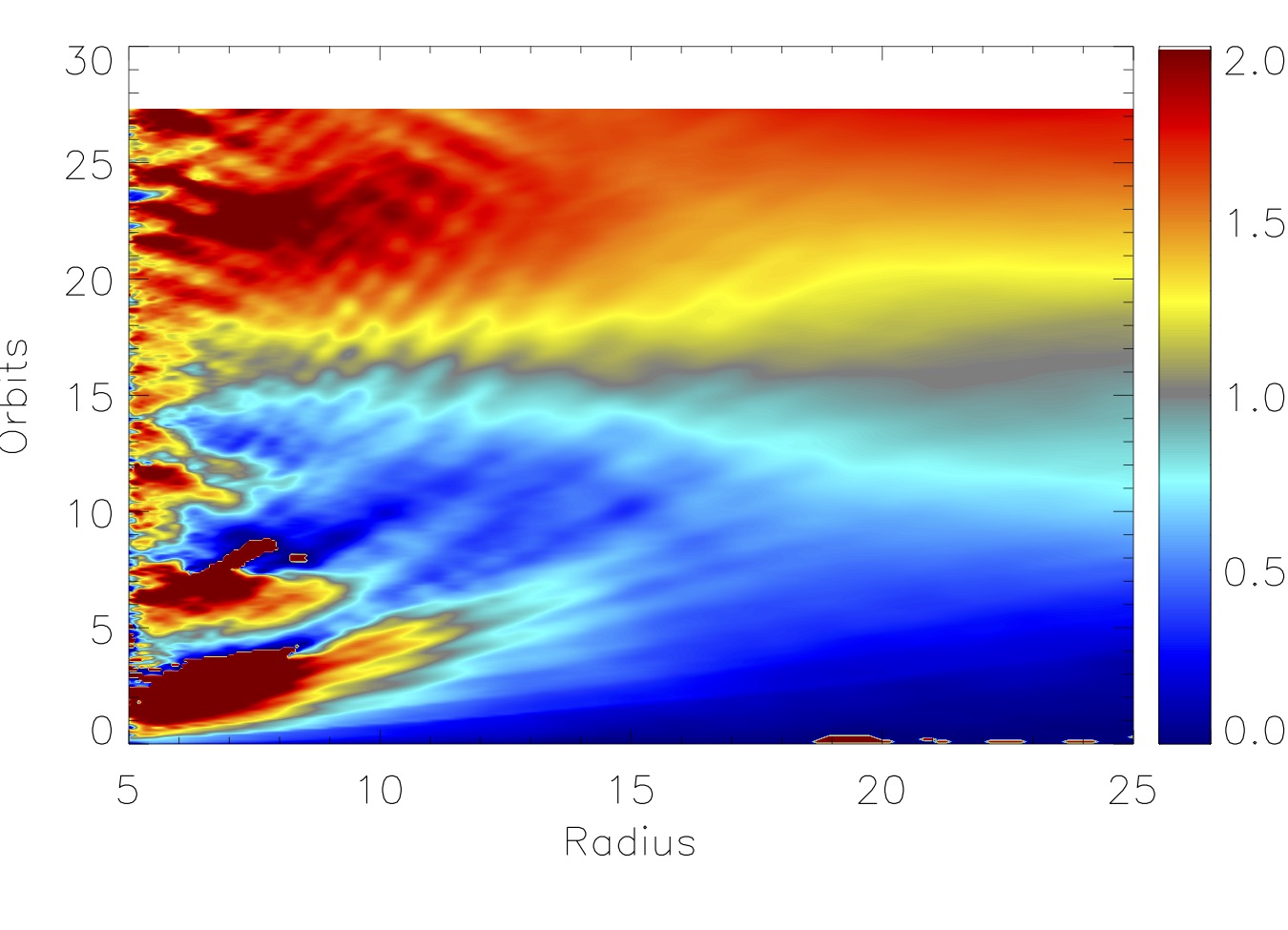}
\includegraphics[width=0.5\textwidth]{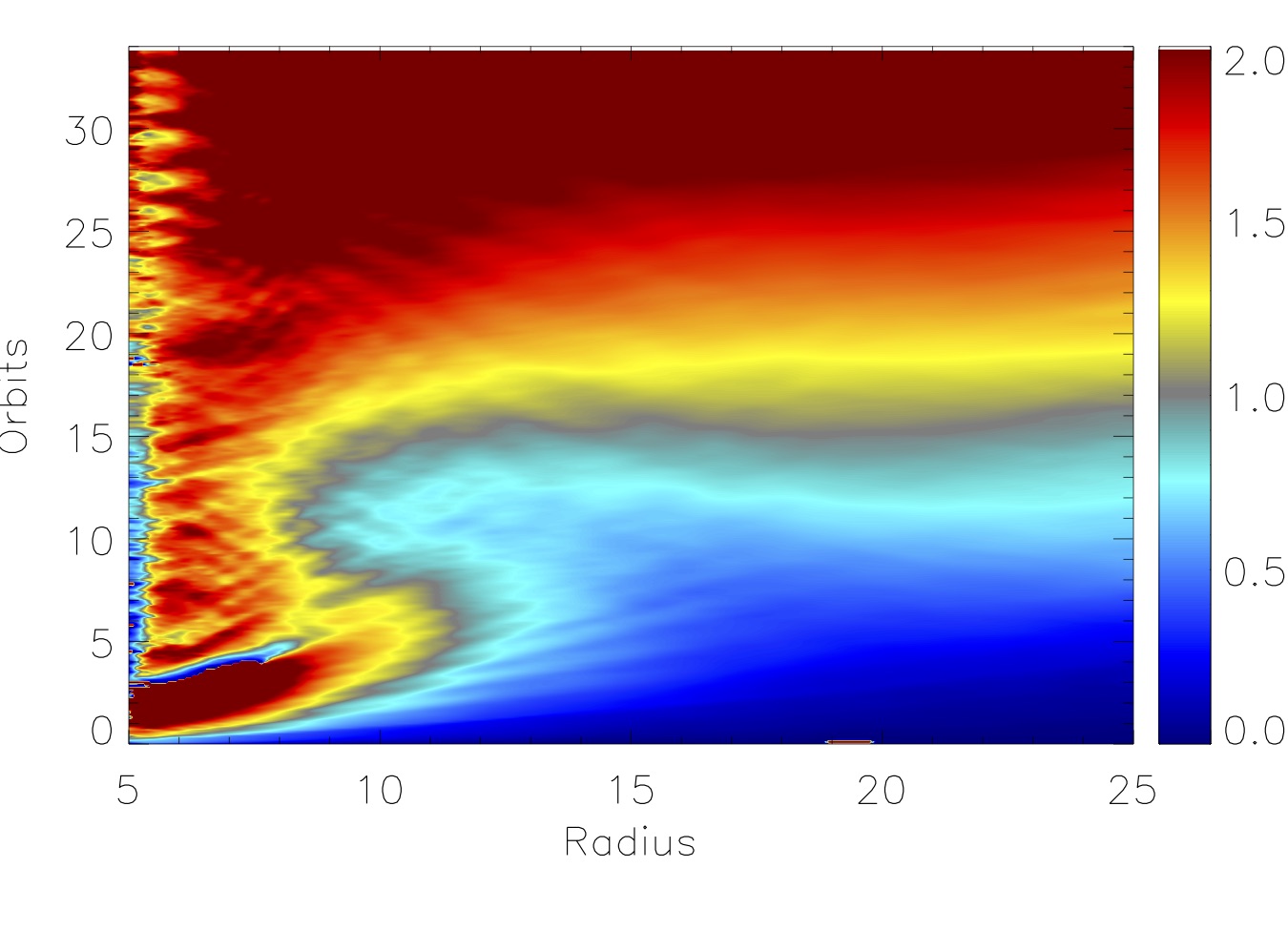}
\includegraphics[width=0.5\textwidth]{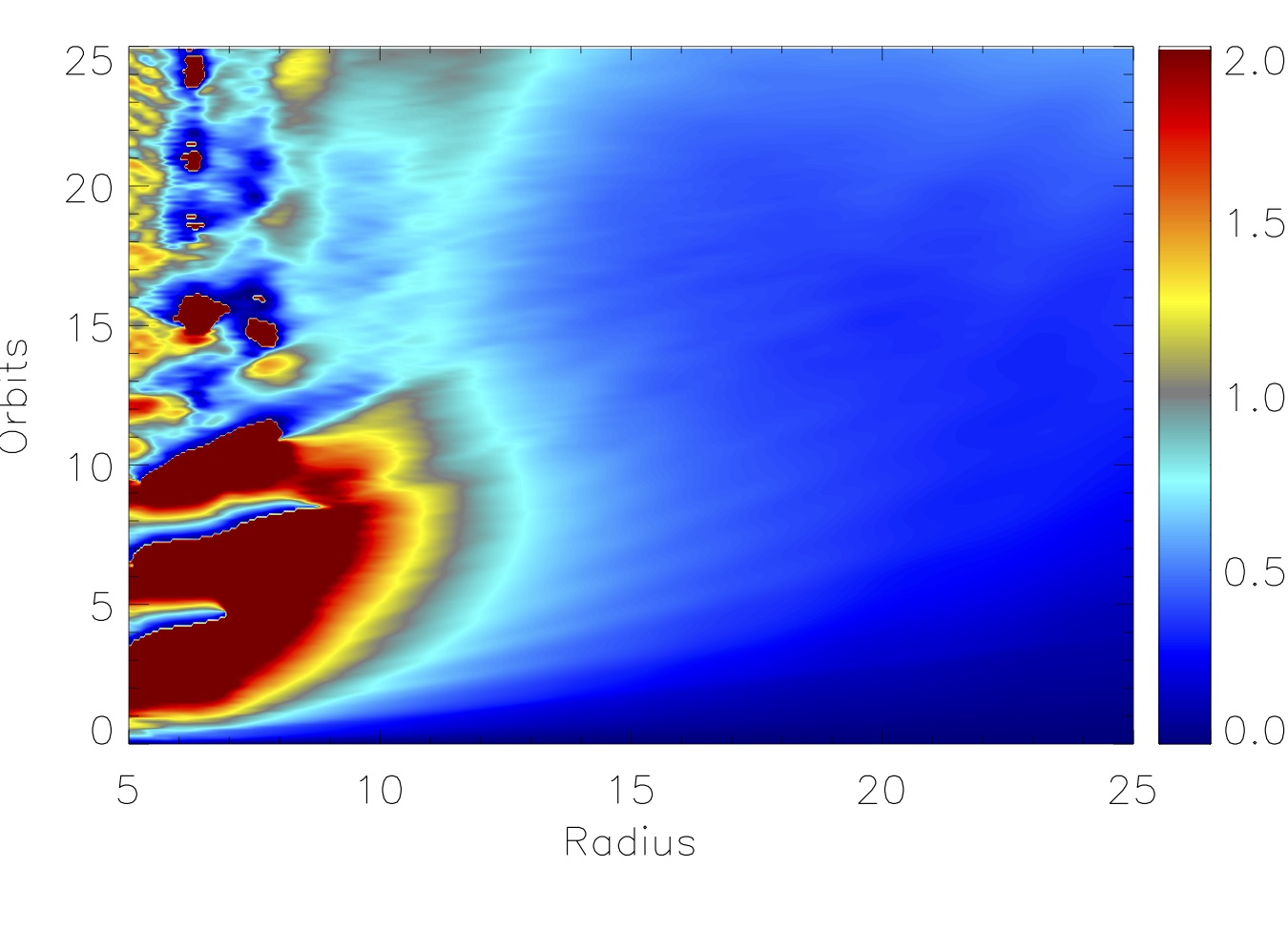}
\caption{Spacetime diagram for the precession angle $\phi$ for the hydrodynamic models Tilt6-H (top), Tilt12-H (middle)  and  Tilt24-H (bottom).  }
\label{fig:phiH}
\end{center}
\end{figure}

Figure~\ref{fig:phiH} shows the spacetime diagrams for the precession angle $\phi$ in the HD models.  All three tilts show initial precession at the Lense--Thirring rate at the fiducial radius, but the rate slows after orbit 5, even reversing in the two lower tilt models.  When the alignment front reaches its maximum radius and stalls and even reverses, the disk outside the front has gone into solid-body precession. In Tilt6-H and Tilt12-H, this happens after orbit 10, and the outer disk precesses at a rate corresponding to the Lense--Thirring frequency at $r=18$.  The outer disk in Tilt24-H approaches solid body precession at the same time, but as the disk has a larger radial extent, its angular momentum-weighted mean radius is greater, and the associated Lense--Thirring precession frequency is considerably smaller.  
The loss of a phase angle gradient  and the onset of solid body precession in the outer disk occurs earlier in the HD models compared to the MHD models.    

\begin{figure}
\begin{center}
\includegraphics[width=0.5\textwidth]{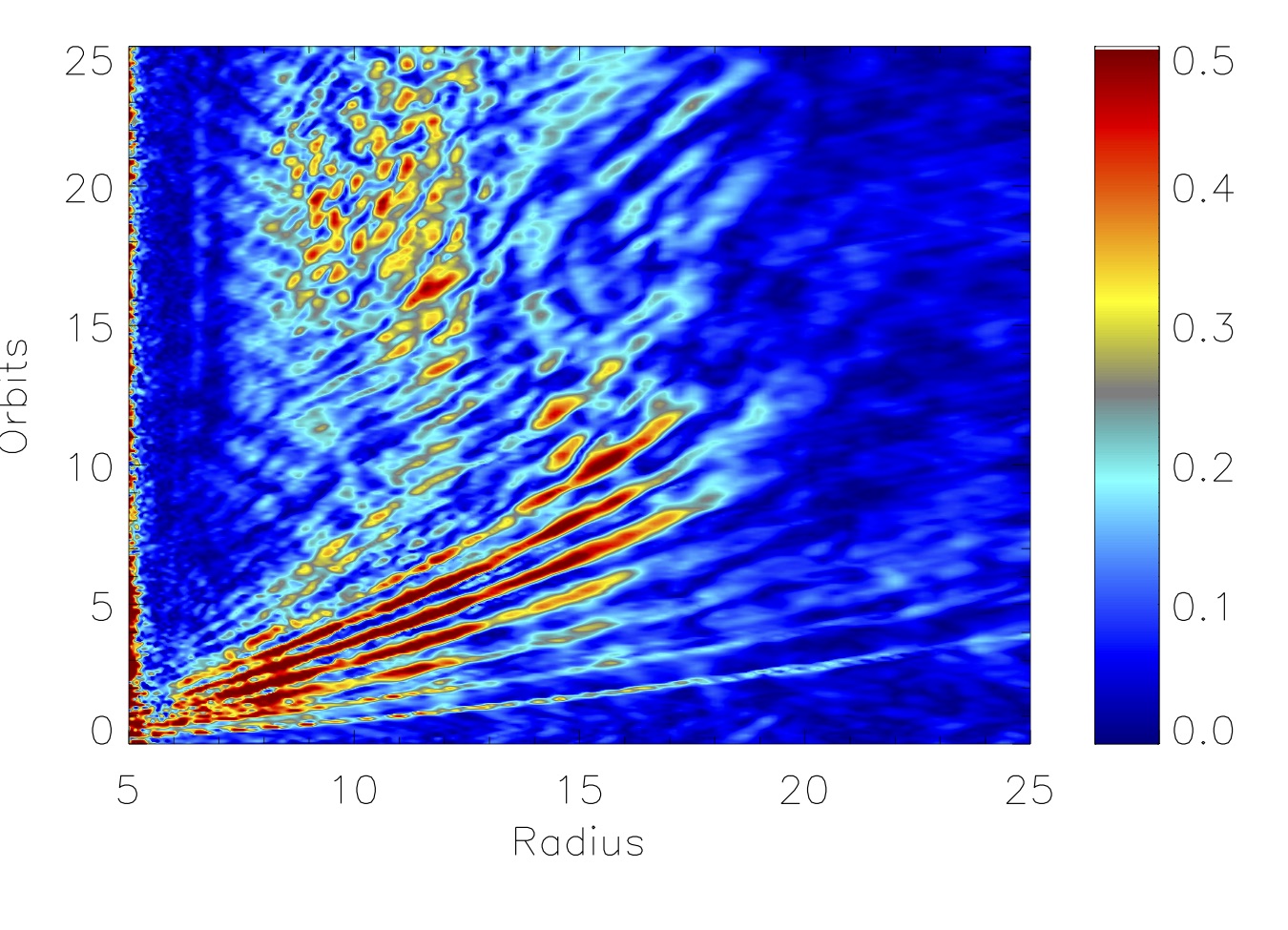}
\includegraphics[width=0.5\textwidth]{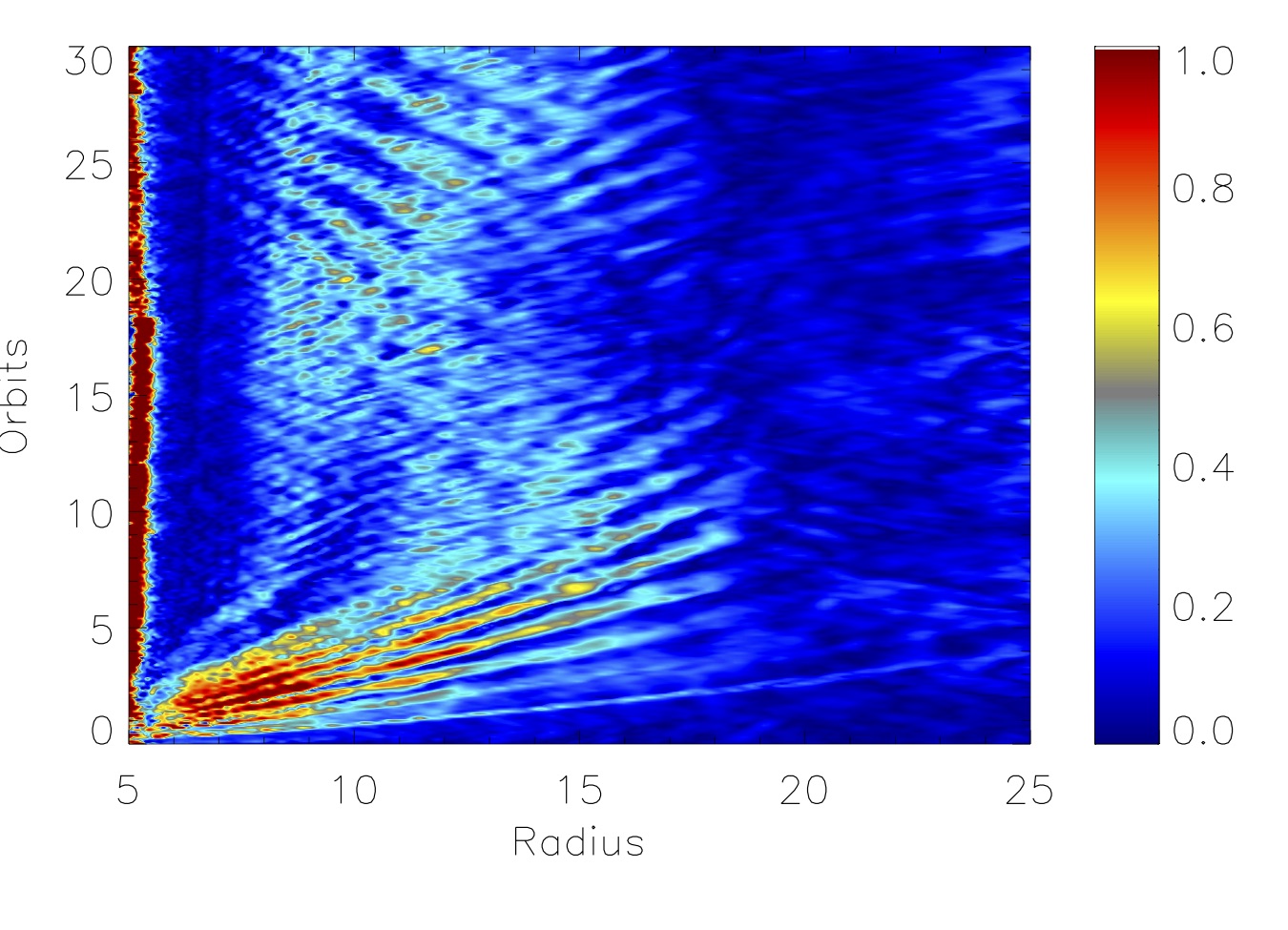}
\includegraphics[width=0.5\textwidth]{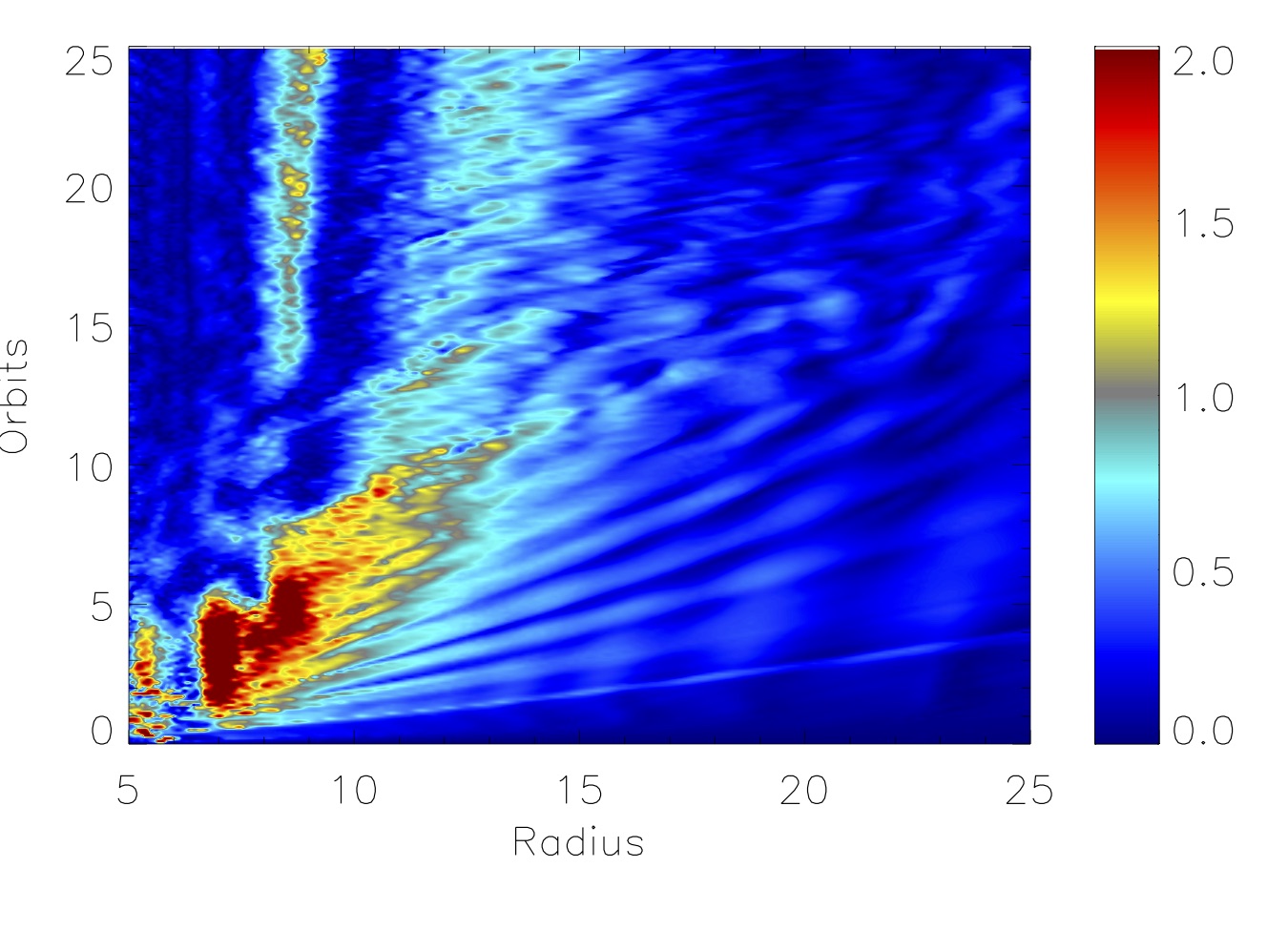}
\caption{Spacetime diagram for the warp $\psi$ (bottom) for the HD models Tilt6-H (top), Tilt12-H (middle)  and  Tilt24-H (bottom). Colors run from $\psi=0$ (blue) to a maximum value (red) of $\psi=0.5$, 1.0 and 2.0 respectively. }
\label{fig:psiH}
\end{center}
\end{figure}

Spacetime diagrams of $\psi$ in the HD models are given in Figure~\ref{fig:psiH}.  As with the MHD figures, the color scale increases proportional to the increase in tilt angle. Again, there is a large initial disk warp in the inner disk in the first 5 orbits.  In sharp contrast to the MHD models, distinct propagating bending waves can be clearly seen propagating throughout the disks in the spacetime plots.  However, the rate at which their amplitudes change with distance and the distinctness of the pulses are sensitive to tilt angle. As all three models have the same sound speed and thickness, $h/r$ is also the same.  Since $\psi$ increases with tilt angle, the normalized $\hat\psi$, which measures the size of the warp with respect to the disk scale height, also increases with tilt.  As a result, bending waves are dissipated more rapidly by shocks in hydro models with greater tilt. It is this effect that accounts for the slower decay of alignment front oscillations in Tilt6-H than in the higher tilt hydro simulations: the bending waves in Tilt6-H are less strongly nonlinear than in the others, permitting them to propagate farther and be more effective in creating local flattening of the precession phase gradient.

The azimuthally averaged surface density profile $\Sigma(r)$ in each of the hydro models shows very little evolution as a function of time, except for that attributable to the presence of waves and some systematic global evolution of the disk with time, e.g., radial spreading due to pressure gradients.  Since there is no internal accretion stress in these models, one would not expect any evolution in $\Sigma(r)$ unless caused by the alignment front itself.  There is no evidence of disk breaking or evolution toward disk breaking.  The surface density $\Sigma (r)$ remains unchanged from its initial value despite the presence of a strong warp at late time.

\section{Conclusions}

We have performed a series of simulations of isothermal disks subject to the Lense--Thirring torque corresponding to three different tilt angles, $\beta_0 = 6^\circ$, $12^\circ$, and $24^\circ$ in both HD and MHD.   By considering a single sound speed and a simple underlying disk model, we can isolate the effect of tilt amplitude on the alignment process. Within the range of $\beta_0$ studied here, tilt has a very limited influence on alignment, especially for MHD disks. To first order, the imposed tilt angle simply sets the ``unit'' and the resulting dynamics are determined by ratios in terms of that unit.  For example, the radial shape of the transition region between the aligned inner disk and unaligned outer disk, as well as the  location of the  head of the alignment front, are very nearly identical for all three models once one scales out $\beta_0$.  This result is consistent with \cite{NP00}  who found that the transition zone between the aligned and unaligned disk spanned the same radial distance for $\beta_0 = 10^\circ$ and $30^\circ$.  Tilt  does have some secondary effects, however.  One such higher order effect is that the steady state warp within the alignment front (defined as $|\partial\hat\ell/\partial\ln r|$) increases at a slightly faster rate with tilt than simple proportionality.  This appears to be due to improved alignment within the inner disk for larger tilts.   Another secondary effect is a small decrease in the initial alignment front velocity with increasing tilt. 

The HD disks behave similarly, but display one additional sensitivity to absolute tilt scale. Because the amplitude of the warp $\psi$ is a function of the tilt angle, for a given disk thickness $h/r$, higher tilt angles have larger $\hat\psi$ values, from which it follows that bending waves are increasingly non-linear, and hence more dissipative.  Bending waves propagating through the disk without hindrance promote solid body precession which, in turn, can end alignment at an earlier time compared to an MHD disk where the turbulence inhibits wave propagation \citep{SKH13b}.  Here, the HD models  see a rapid diminution of precession phase gradient in the outer disk, which brings alignment to a halt and reverses the motion of the alignment front.  Tilt6-H and Tilt12-H exhibit radial oscillations in the location of the alignment front. In contrast, and the continued presence of inward- and out-ward traveling bending waves.   
Tilt24-H model showed no significant front oscillation, did not exhibit significant waves at late time, had a much smaller solid body precession rate in the outer disk, and more closely resembled its MHD counterpart.  Regardless of these effects on the approach to steady-state, the final alignment fronts for HD disks were very similar to one another and to their partner MHD disks once scaled for tilt $\beta_0$.

It has been suggested that large tilts lead to ``disk breaking''.  Some SPH simulations of tilted disks have either seen evidence for a possible separation of the inner and outer disk \citep{Larwood96,PapTerq95,NP00}, or clear disk breaking, whether the central object is a binary or a spinning black hole \citep{Nixon2013, Nealon2015}.   While there are several proposals as to the cause of disk breaking, no single well-supported hypothesis has emerged.  One such conjecture posits that disk breaking occurs at the point where the internal torque from disk viscosity becomes smaller than that due to the Lense--Thirring torque \citep{Nixon2012a}.
Another hypothesis holds that disk breaking occurs where the sound crossing time is comparable to the precession time.  For the soundspeed used here, this is at $r=11.2$.  Both hypotheses suggest that disk breaking should be nearly ubiquitous, occurring relatively near the black hole and for most values of tilt.  

Our simulations find no evidence for disk breaking or even incipient breaking.   For all tilts examined,  the surface density $\Sigma$ remained continuous through the transition region. There was no evidence of a density minimum growing with time.  In the study of \cite{Nixon2012a}, in which viscosity regulated the radial flows,  the development of disk breaking was manifest in the steepening of the scaled gradient of $\beta(r)$ toward a step function; steeper gradients occurred with decreasing internal viscosity.  Here, however, in our MHD models we see only a slightly steeper gradient of $\beta$ with larger tilt,
while for the HD models with no accretion stress, the scaled gradient in $\beta$ through the transition is essentially identical for all the tilts we examined. 

However, the last word on this topic has not yet been said.  Further work that includes inflow-equilibrium surface density profiles, realistic thermodynamics (including both dissipation and cooling), and full relativity lies ahead.

 \section*{Acknowledgements}

This work was partially supported by NSF grants AST-1516299 and AST-1715032 (JHK).  This work was also partially supported by a grant from the Simons Foundation (559794, JHK).
The authors acknowledge the Texas Advanced Computing Center (TACC) at The University of Texas at Austin for providing high performance computing resources that have contributed to the research results reported within this paper.
The National Science Foundation supported this research through XSEDE allocation TG-MCA95C003.  This research is also part of the Blue Waters sustained-petascale computing project, which is supported by the National Science Foundation (awards OCI-0725070 and ACI-1238993) and the state of Illinois. Blue Waters is a joint effort of the University of Illinois at Urbana-Champaign and its National Center for Supercomputing Applications.  The Blue Waters allocation was obtained through an award from the Great Lakes Consortium for Petascale Computation.  Additional simulations were carried out on the {\it Rivanna} high performance cluster at the University of Virginia with support from the Advanced Research Computing Services group.

\bibliography{Bib}

\end{document}